\documentclass[iopscience,apj]{emulateapj}

\usepackage{placeins}
\usepackage{natbib}
\usepackage{amsmath}

\begin{document}

\newcommand\msun{M_{\odot}}
\newcommand\lsun{L_{\odot}}
\newcommand\msunyr{M_{\odot}\,{\rm yr}^{-1}}
\newcommand\be{\begin{equation}}
\newcommand\en{\end{equation}}
\newcommand\cm{\rm cm}
\newcommand\kms{\rm{\, km \, s^{-1}}}
\newcommand\K{\rm K}
\newcommand\etal{{et al}.\ }
\newcommand\sd{\partial}
\newcommand\mdot{\dot{M}}
\newcommand\rsun{R_{\odot}}
\newcommand\yr{\rm yr}

\shorttitle{} 
\shortauthors{Bae et al.}

\title{ACCRETION OUTBURSTS IN SELF-GRAVITATING PROTOPLANETARY DISKS}

\author{Jaehan Bae\altaffilmark{1},
Lee Hartmann\altaffilmark{1},
Zhaohuan Zhu\altaffilmark{2},
Richard P. Nelson\altaffilmark{3}}

\altaffiltext{1}{Dept. of Astronomy, University of Michigan, 500
Church St., Ann Arbor, MI 48105, USA} 
\altaffiltext{2}{Department of Astrophysical Sciences, Princeton University,
4 Ivy Lane, Peyton Hall, Princeton, NJ 08544, USA}
\altaffiltext{3}{Astronomy Unit, Queen Mary University of London, Mile End Road, London E1 4NS, UK}

\email{jaehbae@umich.edu, lhartm@umich.edu, zhuzh@astro.princeton.edu, r.p.nelson@qmul.ac.uk}

\begin{abstract}

We improve on our previous treatments of long-term evolution of protostellar disks by explicitly solving disk self-gravity in two dimensions.
The current model is an extension of the one-dimensional layered accretion disk model of Bae et al.
We find that gravitational instability (GI)-induced spiral density waves heat disks via compressional heating (i.e. $P\rm{d}V$ work), and can trigger accretion outbursts by activating the magnetorotational instability (MRI) in the magnetically inert disk dead-zone.
The GI-induced spiral waves propagate well inside of gravitationally unstable region before they trigger outbursts at $R \lesssim 1$~AU where GI cannot be sustained.
This long-range propagation of waves cannot be reproduced with the previously used local $\alpha$ treatments for GI.
In our standard model where zero dead-zone residual viscosity ($\alpha_{\rm rd}$) is assumed, the GI-induced stress measured at the onset of outbursts is locally as large as $0.01$ in terms of the generic $\alpha$ parameter.
However, as suggested in our previous one-dimensional calculations, we confirm that the presence of a small but finite $\alpha_{\rm rd}$ triggers thermally-driven bursts of accretion instead of the GI + MRI-driven outbursts that are observed when $\alpha_{\rm rd}=0$.
The inclusion of non-zero residual viscosity in the dead-zone decreases the importance of GI soon after mass feeding from the envelope cloud ceases. 
During the infall phase while the central protostar is still embedded, our models stay in a ``quiescent" accretion phase with $\dot{M}_{\rm acc}\sim10^{-8}-10^{-7}~\msunyr$ over $60~\%$ of the time and spend less than $15~\%$ of the infall phase in accretion outbursts.
While our models indicate that episodic mass accretion during protostellar evolution can qualitatively 
help explain the low accretion luminosities seen in most low-mass protostars, detailed tests of the mechanism will require
model calculations for a range of protostellar masses with some constraint on the initial core angular momentum, which affects
the length of time spent in a quasi-steady disk accretion phase.
\end{abstract}

\keywords{accretion disks, stars: formation, stars: pre-main sequence, hydrodynamics, instabilities}

\section{INTRODUCTION}

Recent infrared surveys have shown that the luminosity functions of protostars 
peak near $1 ~L_\odot$, 
and have a significant fraction of objects at sub-solar luminosities \citep[][see also \citealt{dunham14} for review]{enoch09,evans09,kryukova12,dunham13,stutz13}, which seem
too low given the need to accrete the central protostar in typical estimated lifetimes
\citep[e.g,][]{kenyon90}.
After many improvements \citep[e.g.][]{cheng78,terebey84,fatuzzo04,maclow04} to the 
singular isothermal sphere collapse model of \citet{shu77}, theoretical models imply accretion luminosities of $10-100~L_\odot$ for typical mass and radius of low-mass protostars ($0.5~M_\odot$ and $2~R_\odot$).
One plausible solution to this ``luminosity problem'' is that mass infall occurs first to the
disk, and subsequent disk accretion is low for the most of the time, 
with occasional short-lived, rapid accretion outbursts \citep{kenyon90}.
A number of models were developed over decades to explain such episodic accretion events.
Possible mechanisms include thermal instability in the inner disk \citep{bell94}, interactions with companions \citep{bonnell92,pfalzner08,forgan10}, disk fragmentation plus subsequent migration of clumps generated \citep{vb05,vb06,vb10}, and a combination of gravitational instability (GI) and the magnetorotational instability (MRI) \citep{armitage01,zhu09,zhu10a,zhu10b,martin12,bae13a}.

One shortcoming of previous work on GI + MRI-driven outbursts is the use of
simple parameterized $\alpha$ viscosities \citep{shakura73} to represent the mass transport and energy dissipation 
for the GI and the MRI \citep{armitage01,zhu09,zhu10a,zhu10b,martin12,bae13a}.
This allows one to easily evolve models for the long timescales ($\sim$ Myr) needed to follow disk evolution through the infall phase to the T Tauri phase, and to explore a large parameter space as well.
How well the $\alpha$ treatments mimic the nature of GI and the MRI, however, is still controversial.
For example, the intrinsic non-locality of self-gravity can make the appropriateness of an $\alpha_{\rm GI}$ treatment questionable \citep[e.g.][]{balbus99}, although other studies argue that transport via self-gravity is reasonably well described by $\alpha$ parameterizations when the disk is not too massive \citep[e.g.][]{gammie01,lodato04,cossins09,vorobyov10}.

In our previous work \citep[][hereafter Paper I]{bae13a}, adopting $\alpha$ prescriptions to treat the GI and the MRI, we examined disk evolution and outburst behavior in one-dimensional (radial) models.
A layered accretion disk model was implemented in that work, where we solved a separate set of viscous-disk equations in each layer: the magnetically active surface layer (hereafter active layer) and the underlying magnetically inert region (hereafter dead-zone).
We found that outbursts are triggered as the MRI activates in the dead-zone either thermally or through GI, depending on the dead-zone properties.
More specifically, the presence of a small but finite dead-zone residual viscosity generates additional viscous heating in the dead-zone and thus can thermally trigger outbursts starting at or near the inner edge of the disk, instead of the previously found GI + MRI-driven outbursts with zero dead-zone residual viscosity \citep[e.g.][]{zhu10a}.

In this study, we improve the treatment of disk self-gravity by moving to two-dimensional $(R,~\phi)$ models.  
We assume that the disks have a layered structure as in Paper I; we show how this can be accommodated solving only one set of hydrodynamic equations.  
While the overall scenario of accretion outbursts remains valid, the details vary.
We find that, in contrast to a local treatment of GI,
gravitationally unstable regions generate spiral density waves which can propagate into inner disk
regions that are formally GI-stable via the Toomre $Q$ parameter; 
this triggers the MRI at somewhat smaller radii than would be found with $\alpha_{\rm GI}$ treatments.
Also, as we found in Paper I, the presence of a small but finite residual viscosity in the dead-zone 
decreases the importance of GI soon after initial infall phase, 
and is responsible for thermally-driven accretion outbursts instead of GI + MRI-driven bursts with zero $\alpha_{\rm rd}$.
Our results emphasize the importance of following the propagation of waves into innermost disk radii for predicting the resulting accretion luminosity as a function of time and
thus addressing the protostellar luminosity problem.

\section{METHODS}
\label{sec:methods}

\subsection{Basic Equations}
\label{sec:equations}

We use the FARGO-ADSG code \citep{baruteau08} in 2D ($R$, $\phi$) cylindrical coordinates.
In addition to the hydrodynamic equations in the public version we 
add infall, heating sources, and radiative cooling:

\be\label{eqn:mass}
{\partial \Sigma \over \partial t} + \nabla \cdot (\Sigma v) = \dot{\Sigma}_{\rm in}
\en
\be\label{eqn:momentum}
\Sigma \left( {\partial v \over \partial t} + v \cdot \nabla v \right) = - \nabla P - \Sigma \nabla \Phi + \nabla \cdot \Pi + F_{\rm in}
\en
\be\label{eqn:energy}
{\partial E \over \partial t} + \nabla \cdot (Ev) = - P \nabla \cdot v + Q_{\rm +} - Q_{\rm -} + \dot{E}_{\rm in} \,.
\en
In the above equations $\Sigma$ is the surface density, $v$ is the velocity, $P$ is the vertically integrated pressure, $\Phi$ is the gravitational potential including the disk self-gravitational potential, $\Pi$ is the viscous stress tensor, $E$ is the vertically integrated thermal energy per unit area, and $Q_{\rm +}$ and $Q_{\rm -}$ are the total heating and cooling rates, respectively.
The terms $\dot{\Sigma}_{\rm in}$, $F_{\rm in}$, and $\dot{E}_{\rm in}$ indicate the changes in the equations due to the infall model.

Since the main purpose of this paper is to compare the driving of accretion outbursts in
2D with the results from our previous 1D models (Paper I), in the following we compare the equations we solved
to illustrate differences when applicable.

\subsection{Mass Conservation}
\label{sec:mass}

We use the infall model introduced in Paper I, which is based on the model of \citet{cassen81} with modifications: (1) mass flux per unit distance is assumed to be constant over radius in order to avoid a singularity at the centrifugal radius and (2) envelope material does not fall onto the disk inside $20~\%$ of the centrifugal radius in order to mimic the effect of collimated jets and outflows to prevent low angular momentum material from being added to the system.
The basic idea of the infall model comes from an assumption of infall from a uniformly-rotating, spherically symmetric cloud; thus the axial matter has little angular momentum and falls at small radii, while material originally in (near) the equatorial plane has the maximum angular momentum per unit mass and thus defines the instantaneous outer radius of infall to the disk (i.e. the centrifugal radius).
In addition to the modifications, we apply a $10~\%$ $m=2$ density fluctuation to infalling material.
While the $m=2$ perturbation is chosen to consider possible non-axisymmetric infall from a filamentary envelope, we emphasize that the manner perturbations applied is not crucial for generating spiral structures as well as triggering outbursts.
This is because disk rotates fast enough so that the perturbations smear out.
We additionally test with $10~\%$ of random perturbations in each azimuthal grid zone and find no noticeable changes in outcome.
However, it turns out that without any non-axisymmetric perturbations infalling material does not generate asymmetric instabilities/spiral features.
The mass infall rate of the modified model is 
\be
\label{eqn:infall1}
\dot{\Sigma}_{\rm in}(R,t) = {\dot{M}_{\rm in} \over {2\pi R_c (t) R}} \Big[ 1 + 0.1 \cos(2\phi) \Big] ~{\rm if}~0.2R_c \le R \le R_c
\en
and
\be
\label{eqn:infall2}
\dot{\Sigma}_{\rm in}(R,t) = 0~{\rm if}~R<0.2R_c~{\rm or}~R> R_c,
\en
where $R_c (t)$ denotes the centrifugal radius at time $t$ and $\dot{M}_{\rm in}=0.975c_{s}^3 /G$ is the constant total infall mass rate at a given cloud isothermal sound speed for the singular sphere solution \citep{shu77}.
The term $1 + 0.1 \cos(2\phi)$ in Equation (\ref{eqn:infall1}) accounts for the $m=2$ density perturbation in the infall, where $\phi$ is the angle around the rotational axis of the disk.

With this infall model, the radial component of the mass conservation equation becomes
\be
\label{eqn:mass2}
2\pi R {\partial \Sigma \over \partial t} - {\partial \dot{M} \over \partial R} = 2 \pi R \dot{\Sigma}_{\rm in}
\en
where the radial mass flux $\dot{M}$ is defined as $\dot{M} \equiv - 2 \pi R \Sigma v_R$.
Using Equations (\ref{eqn:infall1}) and (\ref{eqn:infall2}), this results in the same form as the mass conservation equation used in Paper I (see their Equation 1).

\subsection{Momentum Conservation}
\label{sec:momentum}

Since infalling material arrives at the disk surface with different radial and azimuthal velocities from those of the disk material, there exists a shear force.
This can be written as $F_{R,{\rm in}}=\dot{\Sigma}_{\rm in}(v_{R,\rm in}-v_{R,\rm disk})$ and $F_{\phi,{\rm in}}=\dot{\Sigma}_{\rm in}(v_{\phi,\rm in}-v_{\phi,\rm disk})$ and added to Equation (\ref{eqn:momentum}), where $v_{R, \rm in}$ and $v_{\phi,\rm in}$ are the velocities of the infalling material (see Equations A7 and A9) and $v_{R, \rm disk}$ and $v_{\phi,\rm disk}$ are the velocities of the disk, respectively.

To facilitate mass and angular momentum transport, we adopt an $\alpha$ disk model \citep{shakura73} where the disk viscosity is calculated as 
\be
\label{eqn:alpha_disk}
\nu = \alpha {c_s^2 \over \Omega}.
\en
Here, $\alpha$ is a dimensionless parameter characterizing the efficiency of mass transport and energy dissipation and $c_s$ and $\Omega$ denote the sound speed and the angular velocity, respectively.
In this study, the $\alpha$ parameter accounts for mass transport and energy dissipation through the MRI ($\alpha_{\rm MRI}$), GI if a non-zero $\alpha_{\rm GI}$
is included in the model, and possible hydrodynamic turbulence in the dead-zone ($\alpha_{\rm rd}$, see below). 

As our simulations evolve the disk as a single layer that represents the full vertical column density of the disk, 
while assuming that the underlying disk model has two layers in the vertical direction (an active layer and a
dead-zone), we introduce an effective viscosity parameter $\alpha_{\rm eff}$ defined as  
\be
\label{eqn:alpha}
\alpha_{{\rm eff}} = {\Sigma_a \alpha_{\rm a} + \Sigma_d \alpha_d\over {\Sigma}},
\en
where $\Sigma_a$ is the surface density of the active layer, $\Sigma_d$ is the surface density of the dead-zone, and $\Sigma = \Sigma_a + \Sigma_d$ is the total surface density.
$\alpha_a$ and $\alpha_d$ are total viscosity parameters in the active layer and the dead-zone, respectively, which are calculated as $\alpha_a = \alpha_{{\rm MRI},a} + \alpha_{{\rm GI},a}$ and $\alpha_d = \alpha_{{\rm MRI},d} + \alpha_{{\rm GI},d} + \alpha_{\rm rd}$.
We explain each term below.

In the model of \citet{gammie96} and later treatments of disk structure, 
the ionization level is not vertically uniform, but varies in a way that 
it decreases toward the disk midplane with a possible sharp transition.
This transition may separate a disk into the magnetically active surface region (i.e. active layer) and the magnetically inert region around the midplane (i.e. dead-zone).
In our fiducial models we assume that the active layer can contain $\Sigma_A = 100~{\rm g~cm^{-2}}$ at maximum via non-thermal ionization \citep{gammie96}.
The MRI viscosity parameter in the active layer ($\alpha_{{\rm MRI},a}$) and the dead-zone ($\alpha_{{\rm MRI},d}$) are assumed to have a fixed value $\alpha_{\rm MRI}=0.01$ only if a region can sustain the MRI.
Thus, $\alpha_{{\rm MRI},a}$ is always set to $\alpha_{\rm MRI}$ by its definition.
On the other hand, $\alpha_{{\rm MRI},d}$ becomes $\alpha_{\rm MRI}$ only if the azimuthally-averaged midplane temperature exceeds the MRI activation temperature $T_{\rm MRI}=1500$~K so that the collisional ionisation of alkali metals (e.g. potassium), or dust sublimation, produces a sufficient ionization level for the dead-zone to thermally sustain the MRI.
Otherwise, $\alpha_{{\rm MRI},d}$ is set to zero.
We use azimuthally-averaged midplane temperatures when activate the MRI in order to be conservative since our treatment for the MRI activation is crude.

As an aside, we note that the idea of an active layer accreting viscously has been challenged by 
\citet{baistone13,bai13,bai14}, who find that the inclusion of ambipolar diffusion limits the effectiveness
of viscous transport, and argue that magnetically-driven winds from upper layers are ultimately
responsible for accretion at radii of order 1 to $10-20$~AU.  
As long as there is some mechanism of mass transport other than GI that results in accretion rates less than the infall rate to the disk, the main features of our models should remain relevant and mass will still pile up to produce outbursts.

We consider cases with either zero or non-zero residual viscosity $\alpha_{\rm rd}$ in the dead-zone.
This is motivated by recent 3D magnetohydrodynamic simulations suggesting that the dead-zone can have some non-zero residual viscosity, which can be as large as $\sim10^{-5}-10^{-3}$, due to hydrodynamic turbulence driven by the Maxwell stress in the active layer \citep{okuzumi11,gressel12}.
In the non-zero $\alpha_{\rm rd}$ case, we use $\alpha_{\rm rd}=10^{-4}$.
We note that the mass accretion rate of the dead-zone cannot exceed that of the active layer ($\dot{M}_d \leq \dot{M}_a$) if the non-zero $\alpha_{\rm rd}$ is due to turbulence propagated from the active layer.
Therefore, we limit $\alpha_{\rm rd}$ as  
\be
\alpha_{\rm rd} = {\rm min} \left( 10^{-4},~\alpha_{\rm MRI} {\Sigma_a \over \Sigma_d}\right).
\en

To isolate the effects of using a local prescription for the GI from the use of 2D vertically-averaged models, we compute some models with an $\alpha_{\rm GI}$ prescription 
\be
\alpha_{\rm GI} = e^{-Q^2},
\en
where $Q \equiv \pi G \Sigma / \Omega c_s$ is the Toomre parameter.
In the models where disk self-gravity is explicitly solved (hereafter self-gravity models), 
$\alpha_{\rm GI}$ is set to zero.

The azimuthal component of the momentum equation becomes 
\be
\label{eqn:momentum2}
2\pi R {\partial \over \partial t}(\Sigma R v_\phi) - {\partial \over \partial R} (\dot{M} R v_\phi) = 2\pi {\partial \over \partial R} (R^2 \Pi_{R\phi})+ 2\pi R^2 \dot{\Sigma}_{\rm in} v_{\phi,{\rm in}}
\en
where we use Equation (\ref{eqn:mass2}) and axisymmetry is assumed. 
If we use $v_\phi = R\Omega$ and the infall model given in Equations (\ref{eqn:infall1}) and (\ref{eqn:infall2}), the momentum equation also has the same form as in Paper I (see their Equation 2).
The only concern here is the viscous stress tensor $\Pi_{R\phi}$ because it has viscosity terms in it that vary between 
the active and dead layers in our underlying model. 
However, if the stress is defined in terms of $\alpha$, one can easily show that $\Pi_{R\phi} = \Pi_{R\phi,a} + \Pi_{R\phi,d}$ by using the effective $\alpha$ parameter introduced in Equation (\ref{eqn:alpha}), assuming the disk is vertically isothermal and the two layers share the same velocity field.
In this case, the momentum equations for the two layers can be added linearly.

\subsection{Energy Conservation}
\label{sec:energy}

We assume that the infalling material has the same temperature as the disk surface (i.e. active layer) at the time of its addition.
Thus, we add the corresponding thermal energy $\dot{E}_{\rm in} = k \dot{\Sigma}_{\rm in} T_a / (\gamma-1)\mu m_{\rm H}$ to the disk where $T_a$ denotes the active layer temperature.
We note that $\dot{E}_{\rm in}$ accounts only for the thermal energy of infalling material.
The heat produced by kinetic energy of infalling material will be discussed below.

The thermal energy of a disk is determined by the balance between total heating and radiative cooling.
Heating includes the internal viscous heating, the external irradiation, the infall heating while it exists, the compressional heating (i.e. $P\rm{d}V$ work), and the artificial viscosity given by the prescription in \citet{vn50}.
The von Neumann-Richtmyer viscosity constant, measuring the number of grid zones over which the artificial viscosity spreads a shock, is set to the default value in FARGO-ADSG code, 1.4.

The viscous heating $Q_{{\rm vis},i}$ is defined as 
\be
Q_{{\rm vis},i} = {1 \over {2\nu_i \Sigma_i}}(\Pi_{RR,i}^2+\Pi_{R\phi,i}^2+\Pi_{\phi\phi,i}^2) + {{2\nu_i \Sigma_i}\over 9} (\nabla \cdot v)^2,
\en
where $\nu_i$ is viscosity calculated as $\nu_i = \alpha_i c_{s}^2/\Omega$ and $\Pi_{RR,i}$, $\Pi_{R\phi,i}$, and $\Pi_{\phi\phi,i}$ are components of the viscous stress tensor.
The subscript $i$ denotes either the active layer (``$a$'') or the dead-zone (``$d$'').
Note that velocity and temperature are assumed to be the same over the two layers while the surface density and viscosity parameter vary when calculating the viscous dissipation.

The external irradiation flux $Q_{\rm irr}$ is the sum of the fluxes from the central star, accretion luminosity, and the envelope:
\be
\label{eqn:heat_ext}
Q_{\rm irr} \equiv \sigma T_{\rm irr}^4 = {f_* L_* \over 4\pi R^2} + {f_{\rm acc} L_{\rm acc} \over 4\pi R^2} + \sigma T_{\rm env}^4.
\en
Here, $T_{\rm irr}$ is the temperature corresponding to the external irradiation flux, $L_*$ and $L_{\rm acc}$ are the stellar and the accretion luminosity, and $T_{\rm env}$ is the envelope temperature. 
The coefficients $f_*$ and $f_{\rm acc}$ account for the non-normal irradiation of the disk
surface and both are set to 0.1 in this study.  
We increase the stellar luminosity as the central star accretes mass, following the mass-luminosity relation 
\be
\label{eqn:Lstar}
\log_{10} \left ({L_* \over \lsun} \right) = 0.20 + 1.74\log_{10} \left( {M_* \over \msun} \right)
\en
which is an approximate power-law fit to the mass-luminosity relation, 
using the luminosities and effective temperatures from \citet{kenyon95} for
Taurus pre-main sequence stars and the \citet{siess00} evolutionary tracks to 
convert the HR diagram positions to masses.
The accretion luminosity is calculated as 
\be
\label{eqn:Lacc}
L_{\rm acc} = {GM_* \dot{M}\over 2 R_{\odot}},
\en
where we assume a typical T Tauri stellar radius of two solar radii. 

During the infall phase kinetic energy carried by the infalling material is dissipated in two ways: immediate shock dissipation at the disk surface and readjustment process within the disk.
While both processes are accompanied by corresponding energy release, the readjustment process, which is due to smaller specific angular momentum of the infalling material than that of the disk material at the same radius, is taken care in the code by adding the proper shear force in the momentum equation as explained in Section \ref{sec:momentum}.  
The shock heating by infalling material (see Appendix for details) corresponding to the infall model outlined in Equations (\ref{eqn:infall1}) and (\ref{eqn:infall2}) is 
\begin{eqnarray}
\label{eqn:q_infall}
Q_{{\rm in}} = {GM_* \dot{M}_{\rm in} \over 4\pi R_c^3} {2-(R/R_c) \over (R/R_c)^2}\Big[ 1 + 0.1 \cos(2\phi) \Big]
\nonumber\\
{\rm if}~0.2R_c \le R \le R_c
\end{eqnarray}
and
\be
Q_{{\rm in}} = 0~{\rm if}~R < 0.2R_c~{\rm or}~R > R_c.
\en
The dissipation of kinetic energy at the shock is treated as an external heating source since it happens near the disk surface \citep{cassen81}.
The infall heating term is thus added at the surface of the disk (see below).

The radiative cooling rate $Q_{-}$ is simply
\be
\label{eqn:cooling}
Q_{-} = 2 \sigma T^4 f(\tau),
\en
where $T$ and $\tau$ are temperature and optical depth at the region where the cooling rate is calculated.
In Equation (\ref{eqn:cooling}), $f(\tau)$ is defined as
\be
\label{eqn:tau}
f(\tau) = {8 \over 3} {\tau \over {1 + \tau^2}},
\en
which is chosen to accommodate both optically thin and thick cooling \citep{johnson03,zhu10a,zhu12}.
The optical depth is calculated as $\tau=\Sigma\kappa/2$ where the Rosseland mean opacity $\kappa$ is taken from \citet{zhu09}.

\begin{deluxetable*}{cccccccccc}
\tablecolumns{13}
\tabletypesize{\tiny}
\tablecaption{Parameters and results\label{tab:results}}
\tablewidth{0pt}
\tablehead{
\colhead{$\alpha_{\rm rd}$} & 
\colhead{$\alpha_{\rm MRI}$} & 
\colhead{$\Sigma_A$} & 
\colhead{$M_*$\tablenotemark{a}} &
\colhead{$M_{\rm disk}$\tablenotemark{a}} &
\colhead{$M_{\rm burst}$\tablenotemark{b}} &
\colhead{$\dot{M}_{\rm max}$\tablenotemark{b}} &
\colhead{$\Delta t_{\rm{burst}}$\tablenotemark{b}} &
\colhead{$D$\tablenotemark{c}} &
\colhead{$D_T$\tablenotemark{d}} \\
\colhead{} & 
\colhead{} & 
\colhead{(${\rm g~cm}^{-2}$)} & 
\colhead{($M_{\odot}$)} &
\colhead{($M_{\odot}$)} &
\colhead{($M_{\odot}$)} &
\colhead{($M_{\odot}~{\rm yr}^{-1}$)} &
\colhead{(yr)} &
 }
\startdata
0 & 0.01 & 100  & 0.76/0.90 & 0.24/0.06 & $1.97\times 10^{-2}$ & $6.19\times 10^{-5}$ & 880 & 0.031 & 0.007   \\
$10^{-4}$ & 0.01& 100 &  0.78/0.93 & 0.22/0.03 & $3.73\times 10^{-3}$ & $1.43\times 10^{-5}$ & 480 & 0.058 & 0.016
\enddata
\tablenotetext{a}{Masses are taken at the end of infall (0.24~Myr) and at the end of calculations (1~Myr).}
\tablenotetext{b}{Outburst quantities are averaged over the T Tauri phase.}
\tablenotetext{c}{Duty cycle for the entire calculation.}
\tablenotetext{d}{Duty cycle during the T Tauri phase.}
\end{deluxetable*}

While the mass and momentum conservation equations can simply be compared to those in Paper I, the comparison of energy equations is more complicated.
The task is to relate the vertically integrated thermal energy per unit area $E$ to the disk midplane temperature.
In order to do this, we first assume that the active layer and the dead-zone has their own vertically-isothermal temperatures $T_a$ and $T_d$.
If only the active layer exists (because either the dead-zone has been enlivened or surface density is low enough), the energy equation simply becomes
\begin{eqnarray}
{\partial E \over \partial t} + \nabla \cdot (E v) & = &- P \nabla \cdot v + Q_{{\rm vis,}a} + Q_{\rm in}f(\tau_a) 
\nonumber\\
& & + 2 \sigma T_{\rm irr}^4 f(\tau_a) - 2\sigma T_a^4 f(\tau_a) + \dot{E}_{\rm in},
\end{eqnarray}
and we can relate the midplane temperature ($T_a$ in this case) to the vertically integrated thermal energy $E$.
Here, $f(\tau)$ is defined as in Equation (\ref{eqn:tau}).

If both active layer and dead-zone exist we can write down an energy equation for each separate layer:
\begin{eqnarray}
\label{eqn:energy_a}
{\partial E_a \over \partial t} + \nabla \cdot (E_a v) & = & - P_a \nabla \cdot v + Q_{{\rm vis,}a} + Q_{\rm in}f(\tau_a) 
\nonumber\\
& & + 2 \sigma T_{\rm irr}^4 f(\tau_a) - 2\sigma T_a^4 f(\tau_a) + 2 \sigma T_d^4 f(\tau_d) 
\nonumber\\
& &  - 2\sigma T_a^4 f(\tau_d)
\end{eqnarray}
and
\begin{eqnarray}
\label{eqn:energy_d}
{\partial E_d \over \partial t} + \nabla \cdot (E_d v) & = & - P_d \nabla \cdot v + Q_{{\rm vis,}d}
\nonumber\\
& &  + 2 \sigma T_a^4 f(\tau_d) - 2\sigma T_d^4 f(\tau_d) + \dot{E}_{\rm in}.
\end{eqnarray}
We note that the above two equations are equivalent to the energy equations used in the layered model of Paper I (see their Equations 12 and 13).
Then, the change in total thermal energy $E$ can be written by adding the two equations, 
\begin{eqnarray}
\label{eqn:energy_tot}
{\partial E \over \partial t} + \nabla \cdot (E v) & = & - P_a \nabla \cdot v - P_d \nabla \cdot v + Q_{{\rm vis,}a} 
\nonumber\\
& & + Q_{{\rm vis,}d} + Q_{\rm in}f(\tau_a) + 2 \sigma T_{\rm irr}^4 f(\tau_a)
\nonumber\\
& &  - 2\sigma T_a^4 f(\tau_a) + \dot{E}_{\rm in}.
\end{eqnarray}

From Equation (\ref{eqn:energy_a}), we can express the term $2\sigma T_a^4$ as 
\begin{eqnarray}
\label{eqn:Ta4}
2\sigma T_a^4 & = & [f(\tau_a)+f(\tau_d)]^{-1} \bigg[- P_a \nabla \cdot v + Q_{{\rm vis},a} + Q_{\rm in} f(\tau_a) 
\nonumber\\
& & + 2\sigma T_{\rm irr}^4 f(\tau_a) + 2\sigma T_d^4 f(\tau_d) - {\partial E_a \over \partial t} - \nabla \cdot (E_a v) \bigg]. 
\end{eqnarray}
Then, by substituting Equation (\ref{eqn:Ta4}) into Equation (\ref{eqn:energy_tot}) we obtain
\begin{eqnarray}
\label{eqn:energy2}
{\partial E \over \partial t} + \nabla \cdot (E v) & = & -{ f(\tau_d) \over {f(\tau_a) + f(\tau_d)}} P_a \nabla \cdot v  - P_d \nabla \cdot v 
\nonumber\\
& & + { f(\tau_d) \over {f(\tau_a) + f(\tau_d)}} Q_{{\rm vis,}a} + Q_{{\rm vis,}d}
\nonumber\\
&  &  + { f(\tau_a) f(\tau_d) \over {f(\tau_a) + f(\tau_d)}} Q_{\rm in} + { f(\tau_a) f(\tau_d) \over {f(\tau_a) + f(\tau_d)}} 2 \sigma T_{\rm irr}^4 
\nonumber\\
& & - { f(\tau_a) f(\tau_d) \over {f(\tau_a) + f(\tau_d)}} 2\sigma T_d^4 + \dot{E}_{\rm in}
\nonumber\\
& & -  { f(\tau_a) \over {f(\tau_a) + f(\tau_d)}} \bigg[ {\partial E_a \over \partial t} + \nabla \cdot (E_a v) \bigg]
\\
& \approx & -{ f(\tau_d) \over {f(\tau_a) + f(\tau_d)}} P_a \nabla \cdot v  - P_d \nabla \cdot v 
\nonumber\\
& & + { f(\tau_d) \over {f(\tau_a) + f(\tau_d)}} Q_{{\rm vis,}a} + Q_{{\rm vis,}d}
\nonumber\\
&  &  + { f(\tau_a) f(\tau_d) \over {f(\tau_a) + f(\tau_d)}} Q_{\rm in} + { f(\tau_a) f(\tau_d) \over {f(\tau_a) + f(\tau_d)}} 2 \sigma T_{\rm irr}^4
\nonumber\\
& &  - { f(\tau_a) f(\tau_d) \over {f(\tau_a) + f(\tau_d)}} 2\sigma T_d^4 + \dot{E}_{\rm in}.
\end{eqnarray}
We find that the last term in Equation (\ref{eqn:energy2})
generally can be neglected in the quiescent state, and is also unimportant 
during outbursts when the thermal energy change is dominated by that in the dead zone.

In the limiting case of $\tau_a, \tau_d \gg 1$, the above equation is simplified to
\begin{eqnarray}
{\partial E \over \partial t} + \nabla \cdot (E v) & = & - {\tau_a \over \tau} P_a \nabla \cdot v - P_d \nabla \cdot v + {\tau_a \over \tau} Q_{{\rm vis,}a} + Q_{{\rm vis,}d} 
\nonumber\\
& &+ {1 \over \tau}  Q_{\rm in}  + {2 \over \tau} \sigma T_{\rm irr}^4 -  {2 \over \tau} \sigma T_d^4 + \dot{E}_{\rm in},
\end{eqnarray}
where $\tau \equiv \tau_a + \tau_d$.

\subsection{Boundary Conditions}
\label{sec:boundary}

A transition is expected in the inner disk from a layered structure to a fully viscous disk
at a radius close enough to the central star that stellar irradiation produces high enough
temperatures so the MRI can be thermally-activated.  
This transition should occur at a smaller radius 
($\sim 0.05 - 0.1$~AU) than our inner boundary $R_{\rm in}=0.2$~AU, 
but taking a smaller inner radius results in excessive computational times.  
We therefore mimic the approximate effect of such a transition by assuming that the disk 
inner boundary is always MRI-active and $\alpha_{{\rm MRI},d}$ varies smoothly over the 
transition region $\Delta R_{\rm trans}=0.1$~AU as 
\begin{eqnarray}
\alpha_{{\rm MRI},d} (R) & = &\alpha_{{\rm MRI},d} (R=R_{\rm in}+\Delta R_{\rm trans}) 
\nonumber\\
& & + \left[ \alpha_{\rm MRI} - \alpha_{{\rm MRI},d} (R=R_{\rm in} + \Delta R_{\rm trans}) \right] \nonumber\\
& &\times  \Big[ 1-\sin \Big({\pi \over 2} {{R-R_{\rm in}} \over\Delta R_{\rm trans}} \Big) \Big]\,.
\end{eqnarray}
We then apply standard open boundary conditions at the inner and outer boundaries:
the radial velocity at the inner boundary is set to be the same as that of the first computation zone if the radial velocity is inward, otherwise it is set to 0 in order to avoid any possible inflow.

\subsection{Initial Conditions and Parameters}
\label{sec:parameters}

We cannot treat the initial collapse phase forming the protostellar core,
so we begin the calculations with a $0.2~\msun$ central protostar, using a small
surrounding disk of mass $0.007~\msun$ with an initial surface density distribution of $\Sigma(R) = 100~(R/{\rm AU})^{-1}~{\rm g~cm^{-2}}$ to avoid
numerical problems (the choice of stellar and disk masses agrees well with those of a recently observed Class 0 protostellar system L1527; \citealt{tobin12}.).
In addition, we assume an $1~\msun$ envelope cloud having uniform angular velocity of $\Omega_c=1.15 \times 10^{-14}~{\rm rad~s^{-1}}$ and temperature of $T_{\rm env}=20$~K.
This yields a net constant infall rate of $\sim3.4 \times 10^{-6}~\msunyr$ for the first $\sim0.24$~Myr of calculations, adding $0.8~\msun$ to the central star + disk in total.
We use inner and outer boundaries of 0.2~AU and 100~AU, with 128 logarithmically spaced radial grid-cells and 128 linearly spaced azimuthal grid-cells.
With this choice, $\Delta R /R$ is constant to 0.05 and grid-cells have comparable radial and azimuthal size at all radii.
We performed short runs with higher numerical resolutions which are restarted at the end of infall phase, and found that the triggering of accretion outbursts is not affected by the resolution.

In the standard model (Section \ref{sec:z_alpha}), we use $\alpha_{\rm MRI} = 0.01$, $\Sigma_A = 100~{\rm g~cm^{-2}}$, and $\alpha_{\rm rd}=0$.
In a companion model (Section \ref{sec:nz_alpha}), we test the effect of non-zero dead-zone residual viscosity with $\alpha_{\rm rd}=10^{-4}$.
Model parameters and outcomes are summarized in Table \ref{tab:results}.

\begin{figure}
\centering
\epsscale{1.2}
\plotone{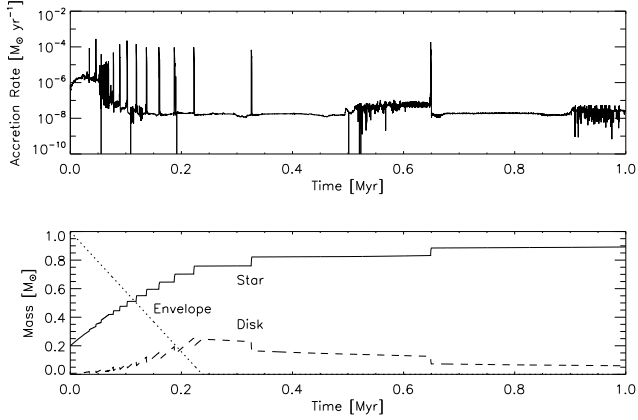}
\caption{(top) Mass accretion rate and (bottom) masses of the central star (solid curve), the disk (dashed curve), and the envelope cloud (dotted curve) as a function of time for the standard $\alpha_{\rm GI}$ model.}
\label{fig:mdot_alpha_z}
\end{figure}

\begin{figure}
\centering
\epsscale{1.2}
\plotone{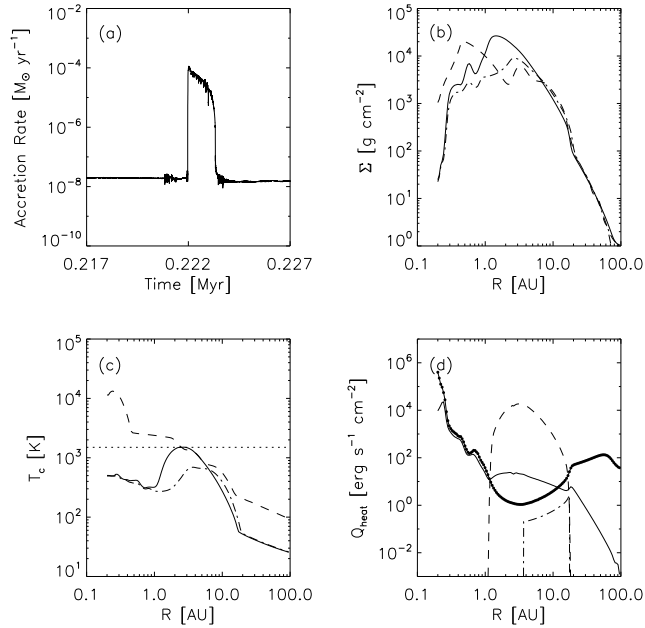}
\caption{(a) Mass accretion rate as a function of time during a single outburst in the standard $\alpha_{\rm GI}$ model. (b) Surface density and (c) midplane temperature distributions at the onset (solid curves), at the peak (dashed curves), and at the end (dash-dotted curves) of the outburst. The horizontal dotted line in panel (c) represents the MRI activation temperature $T_{\rm MRI} = 1500$~K. (d) Contributions of various heating sources at the midplane at the onset of the outburst; external irradiation (solid curve with dots), viscous heating through the MRI (solid curve), GI heating (dashed curve), and infall heating (dash-dotted curve). Radial distributions presented in panels (b) - (d) are taken along the $\phi=0$ direction.}
\label{fig:radial_alpha_z}
\end{figure}

\section{RESULTS}
\label{sec:results}

\subsection{Standard Model ($\alpha_{\rm rd} = 0$)}
\label{sec:z_alpha}

\subsubsection{$\alpha_{\rm GI}$ model}
\label{sec:alpha_model}

We begin with the $\alpha_{\rm GI}$ model.
Figure \ref{fig:mdot_alpha_z} presents the mass accretion rate and masses of the central star, the disk, and the envelope as a function of time.
The overall behavior is similar to that seen in the 1D calculation of Paper I, with outbursts of about
$10^{-4}~\msunyr$ superimposed on a roughly steady accretion rate of $\sim 10^{-6}~\msunyr$ for the first 0.05 Myr, where this background ``quiescent'' rate reduces to $\sim 10^{-8}~\msunyr$ at later times.  
  
\begin{figure}
\centering
\epsscale{1.2}
\plotone{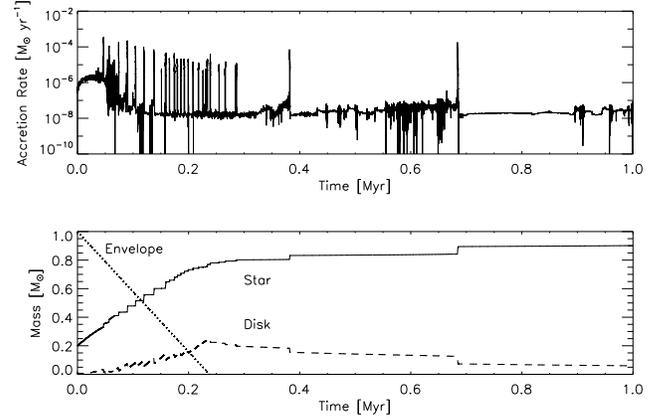}
\caption{(top) Mass accretion rate and (bottom) masses of the central star (solid curve), the disk (dashed curve), and the envelope cloud (dotted curve) as a function of time for the standard self-gravity model. The drops in accretion rate (shown in this figure and other accretion rate plots) are due to the outflow boundary condition adopted and are not physically realistic.}
\label{fig:mdot_z}
\end{figure}

\begin{figure}
\centering
\epsscale{1.2}
\plotone{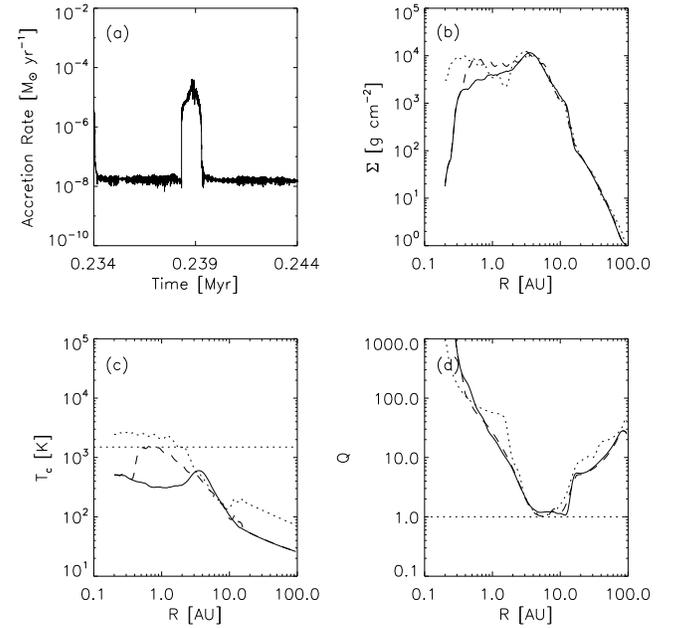}
\caption{(a) Mass accretion rate as a function of time during a single outburst in the standard self-gravity model. (b) Surface density, (c) midplane temperature, and (d) the Toomre $Q$ parameter distributions during quiescent phase (solid curves), at the onset (dashed curves), and at the peak (dotted curves) of the outburst. The horizontal dotted line in panel (c) represents the MRI activation temperature $T_{\rm MRI} = 1500$~K. In panel (d), the dotted line indicates $Q=1$. Radial distributions are taken along the $\phi=0$ direction, but the Toomre Q parameter is azimuthally averaged.}
\label{fig:radial_z}
\end{figure}

\begin{figure*}
\centering
\epsscale{1.1}
\plotone{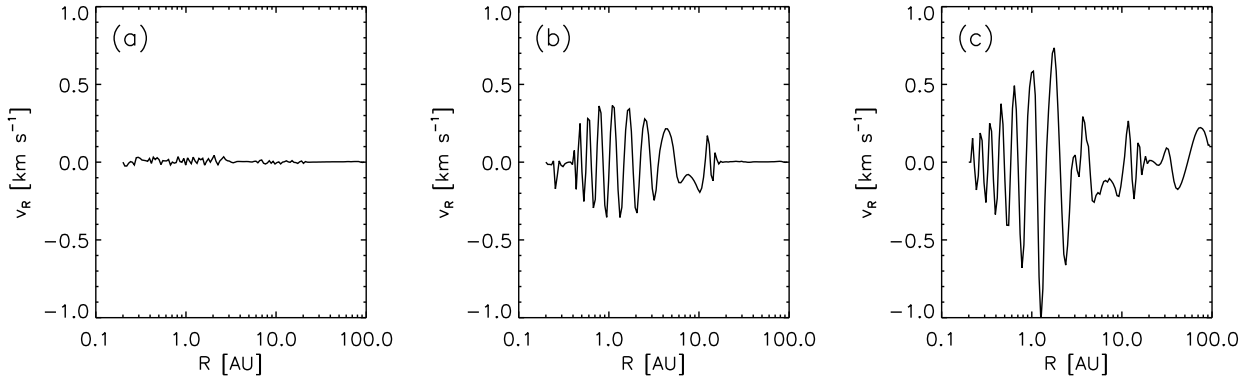}
\caption{Radial velocity profiles along $\phi=0$ (a) during quiescent phase, (b) at the onset and (c) at the peak of the outburst presented in Figure \ref{fig:radial_z}. The velocity profiles show the propagation of GI-induced spiral waves.}
\label{fig:vr}
\end{figure*}

\begin{figure*}
\centering
\epsscale{1.1}
\plotone{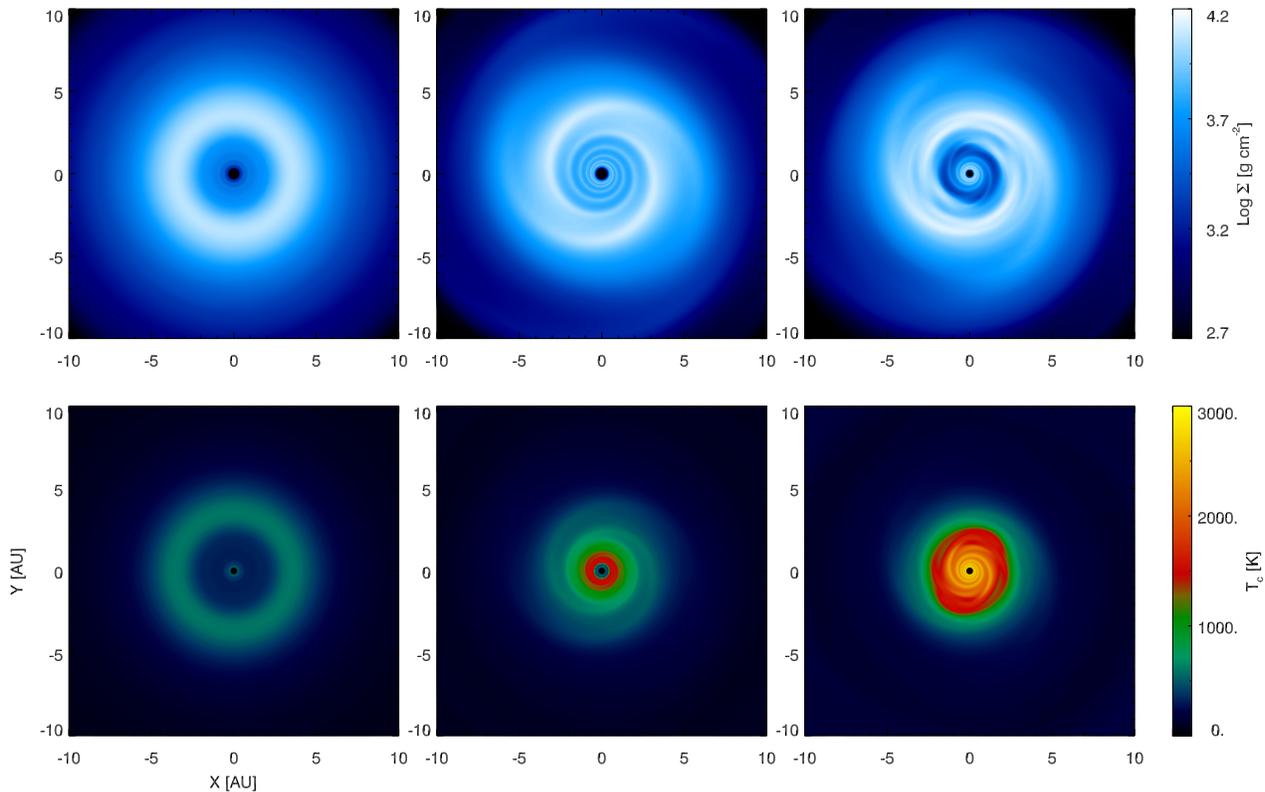}
\caption{(upper) Surface density and (lower) midplane temperature distributions of the inner 10~AU of the disk (left) during quiescent phase before the outburst presented in Figure \ref{fig:radial_z} occurs, and (middle) at the onset and (right) at the peak of the outburst.}
\label{fig:single2d}
\end{figure*}

Looking in more detail at the behavior during an outburst
(Figure \ref{fig:radial_alpha_z}a), the current model does not have such a high initial, short-lived
peak in accretion as in the 1D model.  
This is because
radial pressure gradients were not captured in the 1D calculations which in the 2D case help smooth out the burst.
In addition, the 1D calculations showed a short-lived drop in the mass accretion rate during the main outburst from
$\sim 3 \times 10^{-5}~\msunyr$ to $10^{-6}~\msunyr$ which is not seen in the 2D model.

Figure \ref{fig:radial_alpha_z} illustrates the physical conditions which produce the outbursts, which are
basically the same as in the 1D case. 
Viscous heating through the MRI and external irradiation provide comparable amounts of heating at $R \lesssim 1$~AU.
At $R \gtrsim 20$~AU where disk surface density is low, viscous heating is reduced
while external irradiation dominates.  
At intermediate radii ($1 \lesssim R \lesssim 20$~AU), material piles up due to  
limited mass transport in the dead-zone.
Dissipation by the GI dominates the heating as mass builds up and the outburst is eventually triggered at $\sim2$~AU due to the temperature rise driven by the GI heating.
The MRI-active front then propagates inward, raising the viscosity in the inner disk.
The midplane temperature steeply increases over $10^4$~K at 
the inner $\lesssim 0.5$~AU due to the thermal instability.
These features are essentially the same as in 1D.

We note that since disk self-gravity is not explicitly included in the $\alpha_{\rm GI}$ model, 
no evident spiral structure develops and 
therefore compressional heating and artificial shock heating are negligible at all radii.

\begin{figure*}
\centering
\epsscale{1.1}
\plottwo{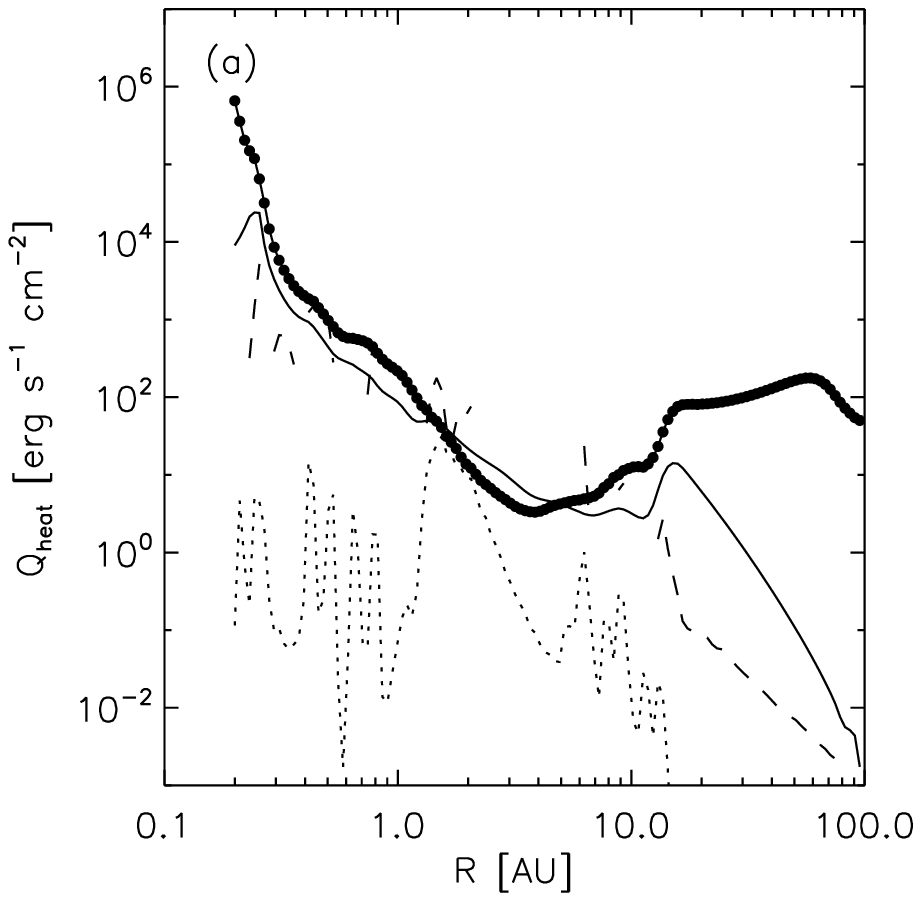}{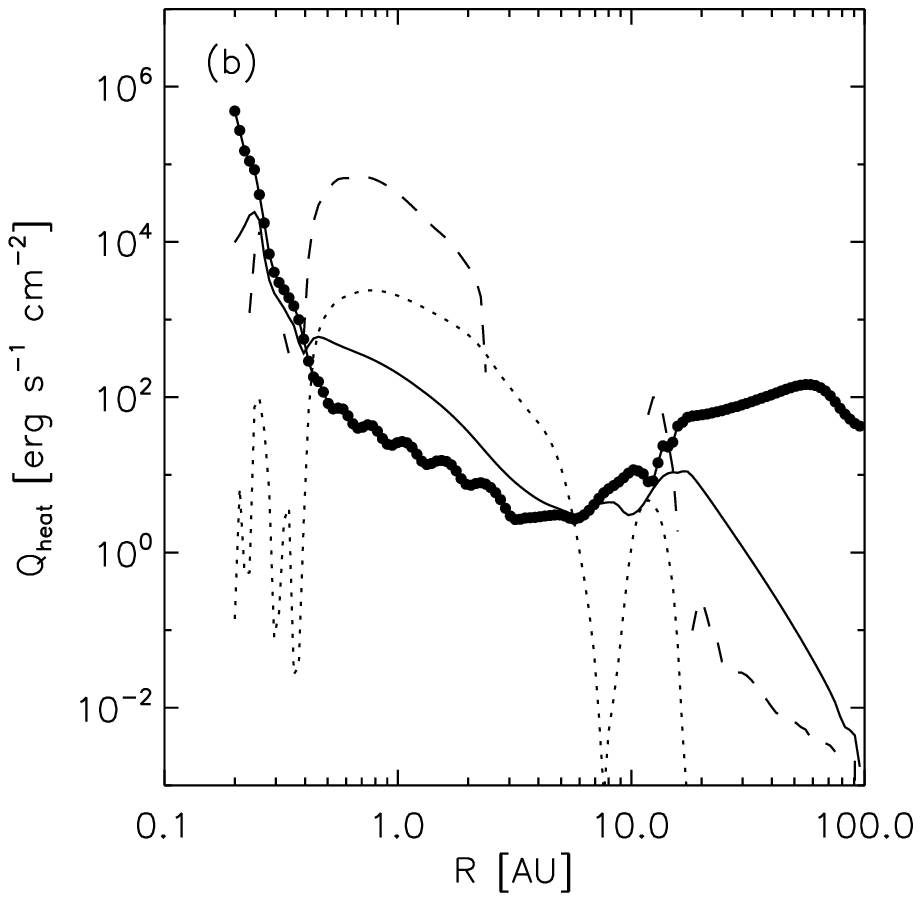}
\caption{Contributions of various heating sources at the midplane (a) during quiescent phase and (b) at the onset of an outburst: external irradiation (solid curve with dots), viscous heating through the MRI (solid curve), compressional heating (dashed curve), and shock dissipation (dotted curve). Compressional and shock dissipation heatings are time-averaged over 1000 years.}
\label{fig:heat_z}
\end{figure*}

\subsubsection{Self-gravity model}
\label{sec:sg_model}

Figure \ref{fig:mdot_z} shows the time evolution of the mass accretion rate and the 
masses of central star, disk, and envelope for the full 2D self-gravity model.
While the overall behavior for the first 0.1~Myr is nearly identical to that of the $\alpha_{\rm GI}$ case,
at later times the self-gravity case exhibits more, smaller bursts of accretion that are
more irregularly-spaced in time.  This is due to the more complex disk structure resulting from
the propagation of spiral waves through the disk. 
The stellar and disk masses at the end of infall phase are $0.76~\msun$ and $0.24~\msun$, which give $M_{\rm disk}/M_*$ of $0.32$.

Figure \ref{fig:radial_z}a shows the mass accretion rate during a single outburst, which 
increases at the beginning of the burst by three orders of magnitude and then gradually increases to $4.1 \times 10^{-5}~\msunyr$ at its peak; the outburst lasts for $1000$~years over which time a 
total mass of $0.01~\msun$ is accreted.
The burst is about a factor of 2-3 lower in peak accretion rate than the $\alpha_{\rm GI}$ model, lasts about 2/3
as long, and exhibits a more ``rounded'' form.  These differences can be traced to differences in the way
the outburst is triggered.  
As shown in Figure \ref{fig:radial_z}, in the self-gravity case the outburst is triggered at smaller radii and at smaller surface
densities, which result in a weaker and shorter accretion episode.  
The lower maximum accretion rate also results in a failure to trigger the thermal instability, which in turn does not produce the very sharp initial peak in mass accretion seen in Figure 2.  

The outburst is triggered differently in the self-gravity case by the propagation of spiral waves into inner disk regions which are formally GI-stable (Figure \ref{fig:vr}).
The velocity perturbations of order $0.5~\kms$ propagate inward and trigger thermal activation of the MRI.  
Two-dimensional distributions of surface density and midplane temperature before, at the onset, and at the peak of the outburst are presented in Figure \ref{fig:single2d}, which also show the propagation of spiral density waves and consequent outburst triggering.
Thus, the essentially non-local aspect of GI produces a quantitative difference in the behavior of the
outburst.

Figure \ref{fig:heat_z} presents contributions of heating sources during the quiescent phase and at the onset of an outburst.
During the quiescent phase, external irradiation and viscous heating via the MRI provide comparable amounts of heat, and dominate disk heating at all radii but the outer disk ($R \gtrsim 10$~AU) where external irradiation dominates.
The disk is gravitationally stable during the quiescent phase, and thus compressional heating through $P{\rm d}V$ work and shock dissipation are less important than other heating sources.
As the disk becomes gravitationally unstable, spiral density waves are generated accompanying a rapid inward accretion at inner disk.
In this example, the inward radial velocity peaks at $\sim2$~AU inside of which radii the compressional heating dominates ($dv_R/dR < 0$).
We emphasize that $P {\rm d}V$ work is the dominating heating source at the radii providing orders of magnitude greater heat than viscous heating and external irradiation heating.
It is also worth to note that rarefactional cooling occurs at $\sim2-10$~AU because $dv_R/dR > 0$ over the region.

\subsection{Effect of Non-zero Residual Viscosity in the Dead-Zone ($\alpha_{\rm rd} = 10^{-4}$)}
\label{sec:nz_alpha}

\subsubsection{$\alpha_{\rm GI}$ model}

\begin{figure}
\centering
\epsscale{1.2}
\plotone{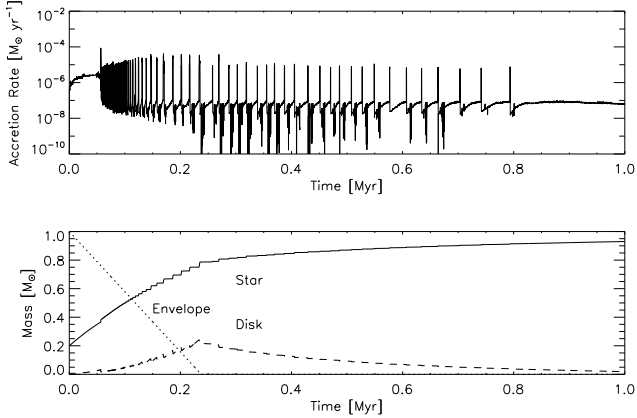}
\caption{(top) Mass accretion rate and (bottom) masses of the central star (solid curve), the disk (dashed curve), and the envelope cloud (dotted curve) as a function of time for the $\alpha_{\rm GI}$ model with $\alpha_{\rm rd}=10^{-4}$.}
\label{fig:mdot_alpha_nz}
\end{figure}

Figure \ref{fig:mdot_alpha_nz} shows the mass accretion rate and the masses of the central star, the disk, and the envelope cloud as a function of time.
As in the standard model, the overall evolution shows a qualitative resemblance to the $\alpha_{\rm GI}$ model in one-dimension (c.f. Figure 6 in Paper I).
However, we note that the outbursts have higher peaks than in the 1D case, which results in faster depletion of the disk.

\begin{figure}
\centering
\epsscale{1.2}
\plotone{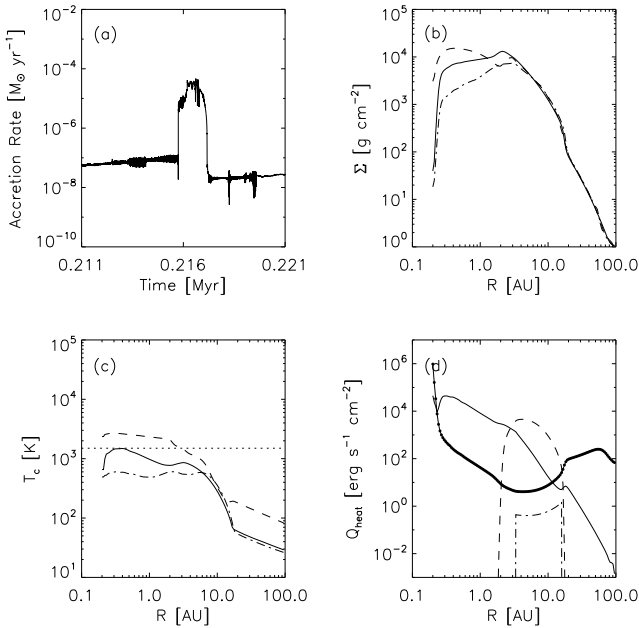}
\caption{Same as Figure \ref{fig:radial_alpha_z} but for the $\alpha_{\rm GI}$ model with $\alpha_{\rm rd}=10^{-4}$. In panel (d), the solid curve includes viscous heating through the MRI and hydrodynamic turbulence in the dead-zone (i.e. non-zero $\alpha_{\rm rd}$) as well.}
\label{fig:radial_alpha_nz}
\end{figure}

To compare outburst behaviors we plot the mass accretion rate during a single outburst in Figure \ref{fig:radial_alpha_nz}.
Radial profiles of surface densities and midplane temperatures at the onset, peak and end of the outburst, as well as contributions of various heating sources to the 
midplane temperature are also plotted in the same figure.
In the non-zero $\alpha_{\rm rd}$ model, the dead-zone residual viscosity generates a significant amount of heating which dominates at $R \lesssim 3$~AU.
It is greater than the external irradiation over these radii by as much as two orders of magnitude.
GI heating is significant at $2 \lesssim R \lesssim20$~AU due to large mass in the dead-zone, but outbursts are thermally triggered near the disk inner edge before enough material piles up for GI to initiate outbursts.

\begin{figure}
\centering
\epsscale{1.2}
\plotone{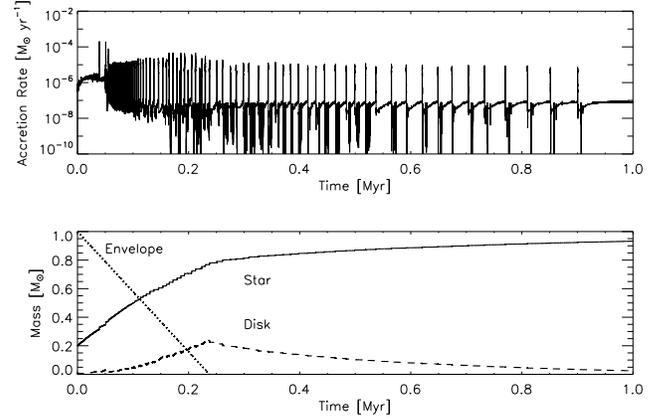}
\caption{(top) Mass accretion rate and (bottom) masses of the central star (solid curve), the disk (dashed curve), and the envelope cloud (dotted curve) as a function of time for the self-gravity model with $\alpha_{\rm rd}=10^{-4}$.}
\label{fig:mdot_nz}
\end{figure}

\begin{figure}
\centering
\epsscale{1.2}
\plotone{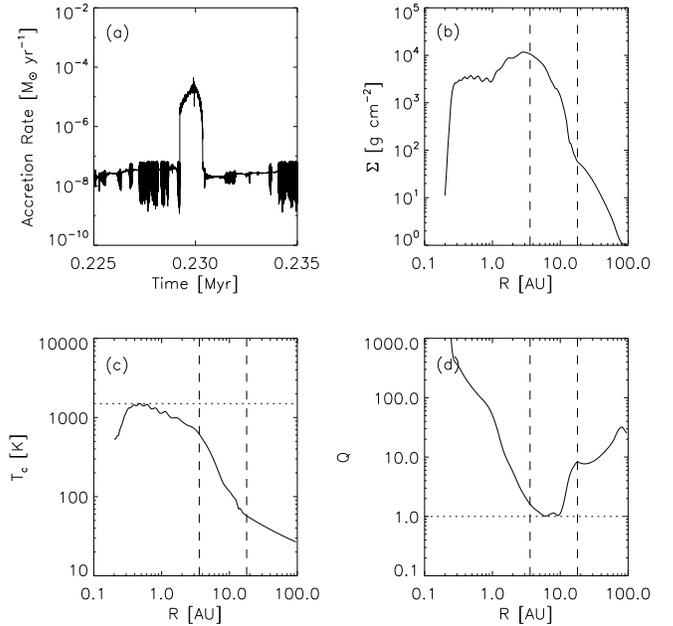}
\caption{(a) Mass accretion rate as a function of time for an outburst occurred during the infall phase ($t \sim 0.23$~Myr) when disk self-gravity is important. Radial distributions of (b) surface density, (c) midplane temperature, and (d) the Toomre $Q$ parameter at the beginning of the outburst are plotted as well. Horizontal dotted line in panel (c) indicates the MRI activation temperature $T_{\rm MRI}$ and the one in panel (d) shows where $Q=1$. The vertical dashed lines present the radii between which infalling material from the envelop cloud falls on at this time. Radial distributions are taken along the $\phi=0$ direction, but the Toomre Q parameter is azimuthally averaged.}
\label{fig:single_nz1}
\end{figure}
\begin{figure}
\centering
\epsscale{1.2}
\plotone{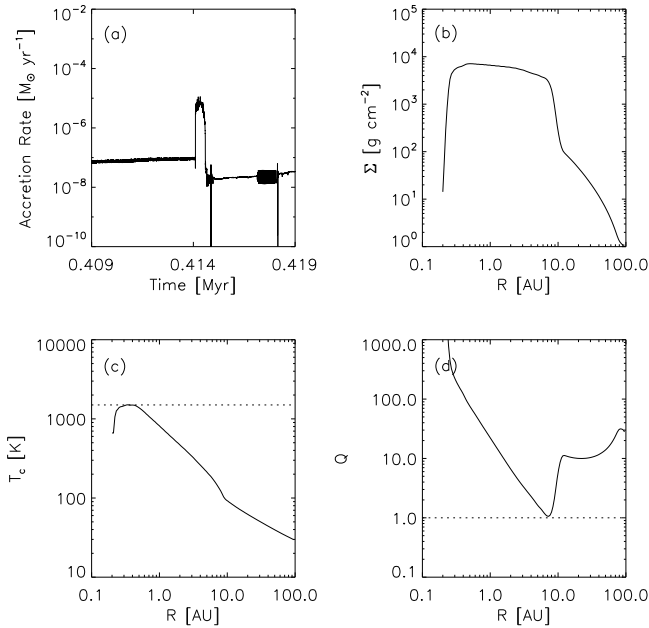}
\caption{Same as Figure \ref{fig:single_nz1} but for an outburst that occurred during the ``T Tauri phase" ($t \sim 0.41$~Myr after infall has stopped) when disk self-gravity becomes negligible. We note that the whole disk is gravitationally stable ($Q > 1$) and there is no signature of spiral waves propagating in the surface density and midplane temperature distributions.}
\label{fig:single_nz2}
\end{figure}

\subsubsection{Self-gravity model}

In Figure \ref{fig:mdot_nz}, we plot the mass accretion rate for the non-zero $\alpha_{\rm rd}$ model as a function of time.
The mass accretion rate maintains a value of $10^{-8}-10^{-7}~\msunyr$ in between bursts, which is in agreement with the zero $\alpha_{\rm rd}$ model, but the outbursts generally have a smaller peak accretion rate $\sim10^{-6}-10^{-5}~\msunyr$ than the ones in the zero $\alpha_{\rm rd}$ model.
At the end of infall phase, stellar and disk masses are $0.78~\msun$ and $0.22~\msun$ giving $M_{\rm disk}/M_*$ of $0.28$.

\begin{figure}
\centering
\epsscale{1}
\plotone{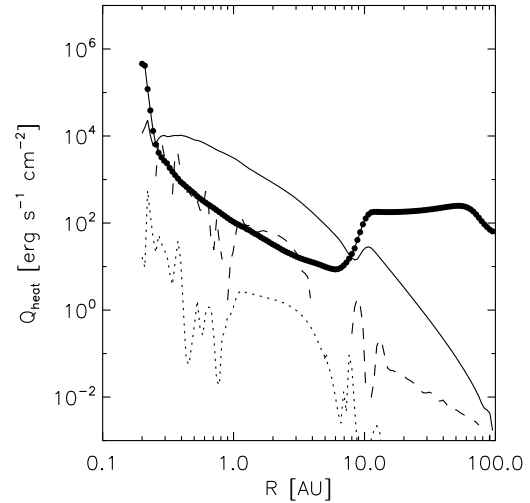}
\caption{
Contributions of various heating sources at the midplane at the onset of the thermally-driven outburst presented in Figure \ref{fig:single_nz2}: external irradiation (solid curve with dots), viscous heating through the MRI plus dead-zone residual viscosity (solid curve), compressional heating (dashed curve), and shock dissipation (dotted curve). Compressional and shock dissipation heatings are time-averaged over 1000 years. Note that the internal viscous heating dominates at $R \lesssim 10$~AU with the help of non-zero $\alpha_{\rm rd}$.}
\label{fig:heat_nz}
\end{figure}

Figure \ref{fig:single_nz1} shows the accretion rate of an outburst that occurred during the infall phase and the radial profiles of surface density, midplane temperature, and the Toomre $Q$ parameter at the beginning of the outburst.
During the infall phase when the disk is fed by infalling material, the outburst-driving mechanism is similar to that of the standard model: spiral density waves propagate inward starting from the gravitationally-unstable outer disk, triggering the MRI in the dead-zone through compressional heating.
However, after infall stops the inner disk is viscously heated and thermally-driven bursts are triggered before material piles up at larger radii.
The transition between the GI + MRI-driven outbursts and the thermally-driven outbursts occurs soon after the mass feeding from the envelop cloud is ceased, at $t \sim 0.3$~Myr.
Figure \ref{fig:single_nz2} shows the accretion rate and radial profiles of surface density, midplane temperature, and the Toomre $Q$ parameter at the initiation of an outburst occurring after the infall phase.
As shown, the outburst is thermally triggered near the disk inner edge before the outer disk becomes gravitationally unstable.
We note that there is no signature of spiral waves at the initiation of the burst.
Contributions from various heating sources at the onset of thermally-driven outburst are plotted in Figure \ref{fig:heat_nz}.
As seen, the inner disk ($R \lesssim 10$~AU) is mainly heated by viscous heating in the dead-zone and heating from $P\rm{d}V$ work and shock dissipation is less important.

\section{DISCUSSION}

\subsection{GI-induced Spiral Density Waves}
\label{sec:diss_transport}

\begin{figure*}
\centering
\epsscale{1.1}
\plotone{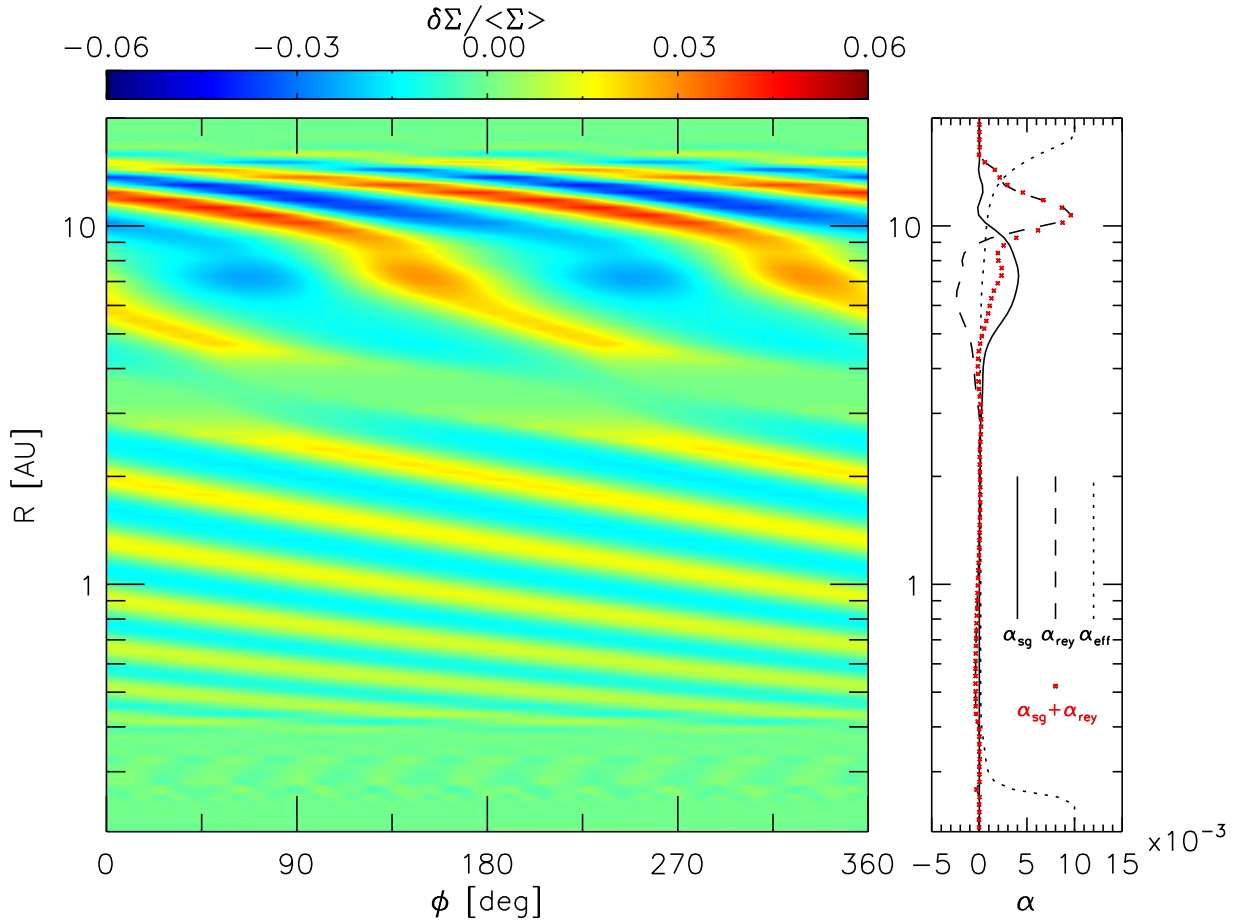}
\caption{(left) Spatial distribution of surface density enhancement/deficit $\delta \Sigma / \langle \Sigma \rangle$ on the $\phi-\log R$ plane at the onset of the outburst presented in Figure \ref{fig:radial_z}. (right) Azimuthally-averaged radial profiles of $\alpha_{\rm sg}$, $\alpha_{\rm rey}$, and $\alpha_{\rm eff}$ are plotted. The total stress induced by GI ($\alpha_{\rm sg} + \alpha_{\rm rey}$) is also plotted with red crosses.}
\label{fig:sdw}
\end{figure*}

As we have described, the propagation of GI-induced spiral density waves plays a crucial role in triggering accretion outbursts 
and thus in the evolution of protoplanetary disks.
Figure \ref{fig:sdw} illustrates the spatial distribution of perturbations to the surface density $\delta \Sigma / \langle \Sigma \rangle$ in the 
$\phi - \log R$ plane at the onset of the GI + MRI-driven outburst presented in Figure \ref{fig:radial_z}. 
As seen in the figure, $m=2$ trailing spiral density waves are dominant.
They originate at $\sim7$~AU where the disk is gravitationally most unstable,
while extending over a range of disk radii from $\sim0.4$~AU to $\sim15$~AU.

In order to measure the strength of the GI-induced stress, we calculate the gravitational shear stress in terms of an effective $\alpha$ \citep{lynden-bell72,gammie01} as 
\be
\alpha_{\rm sg} = - \left( {{d \ln \Omega} \over {d \ln R}} \right)^{-1} {\langle \int_{-\infty}^{\infty} {g_R g_\phi / (4\pi G)} dz \rangle \over \langle \Sigma c_s^2 \rangle}, 
\en
where $g_R$ and $g_\phi$ are self-gravitating acceleration in $R$ and $\phi$ directions and the brackets denote the azimuthal average.
The vertical integration in the above equation is numerically done in the FARGO-ADSG code by changing $B^2$ to $B^2 + \eta^2$ in equations (A1) and (A3) of \citet{baruteau08}, where $\eta$ is defined as $z = \eta R$ (see Appendix A of \citealt{baruteau11}).
We vary $\eta$ evenly by 0.01 from 0 to 1 for the integration (C. Baruteau 2014, private communication).
In addition to the stress directly generated from the gravitational field, GI also produces density and velocity fluctuations that contribute to mass transport and heat dissipation.
This can be quantified using the Reynolds stress calculated as
\be
\alpha_{\rm rey} = - \left( {{d \ln \Omega} \over {d \ln R}} \right)^{-1} {{\langle \Sigma \delta v_R \delta v_\phi \rangle} \over \langle \Sigma c_s^2 \rangle},
\en
where $\delta v_R = v_R - \langle v_R \rangle$ and $\delta v_\phi = v_\phi - \langle v_\phi \rangle$.

The azimuthally-averaged radial profiles of $\alpha_{\rm sg}$ and $\alpha_{\rm rey}$ are plotted on the right panel of Figure \ref{fig:sdw}.
At the initiation of the outburst, gravitational stress $\alpha_{\rm sg}$ is 0.004 at the radius where the spiral waves are generated.
However, the GI-induced spiral waves generate additional hydrodynamic turbulence across a broader region.
In terms of $\alpha_{\rm rey}$, the stress is as large as 0.01 at $\sim10$~AU.
Also, we note that while gravitationally stable at $R \lesssim 1$~AU the propagating spiral waves provide $\sim10^{-3}$ of $\alpha_{\rm rey}$ in the region.
We note that the mass transport through the MRI across this inner region is limited ($\alpha_{\rm eff}\sim10^{-4}$) because of relatively large mass in the dead-zone.

\begin{figure*}
\centering
\epsscale{1.1}
\plottwo{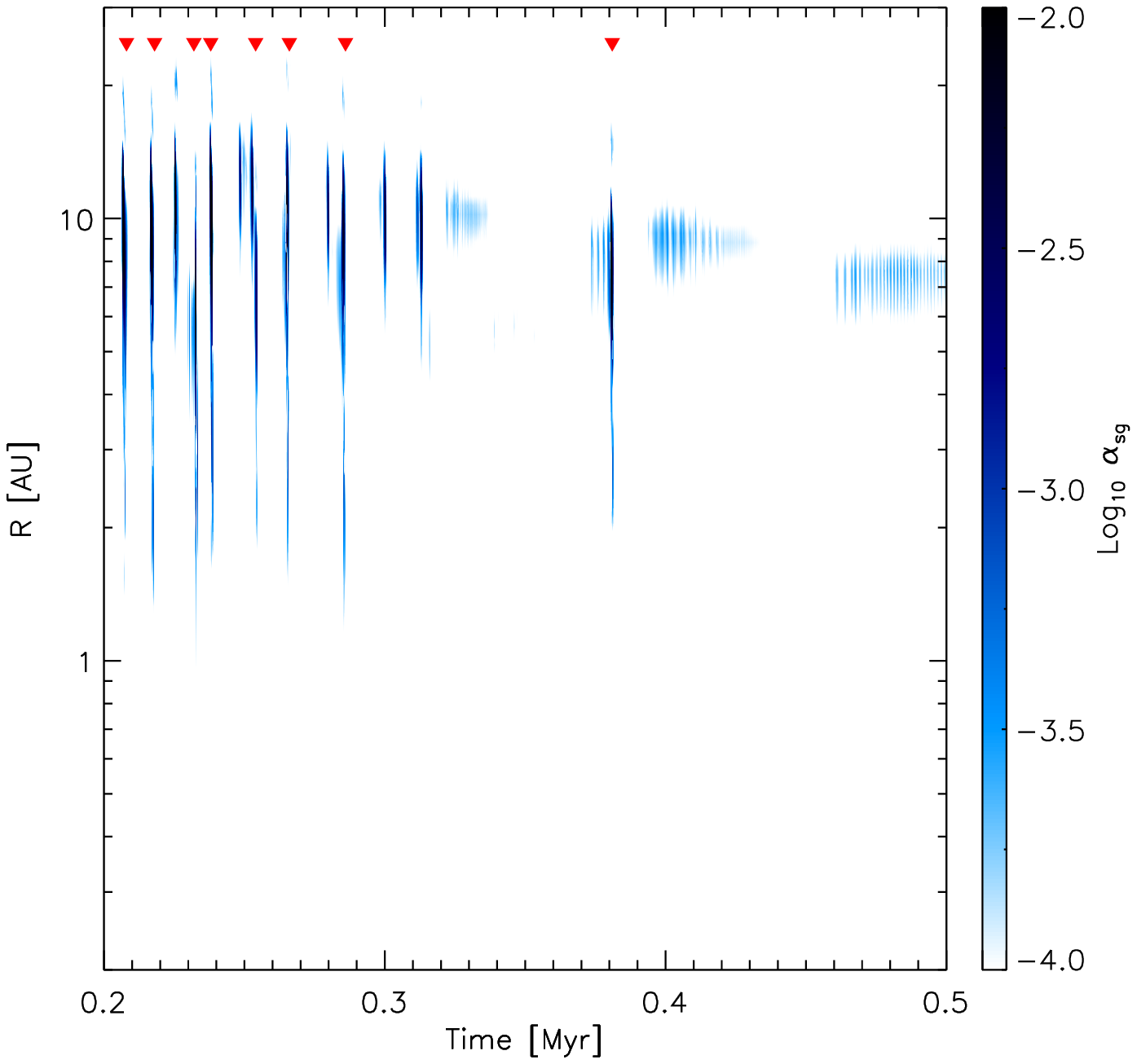}{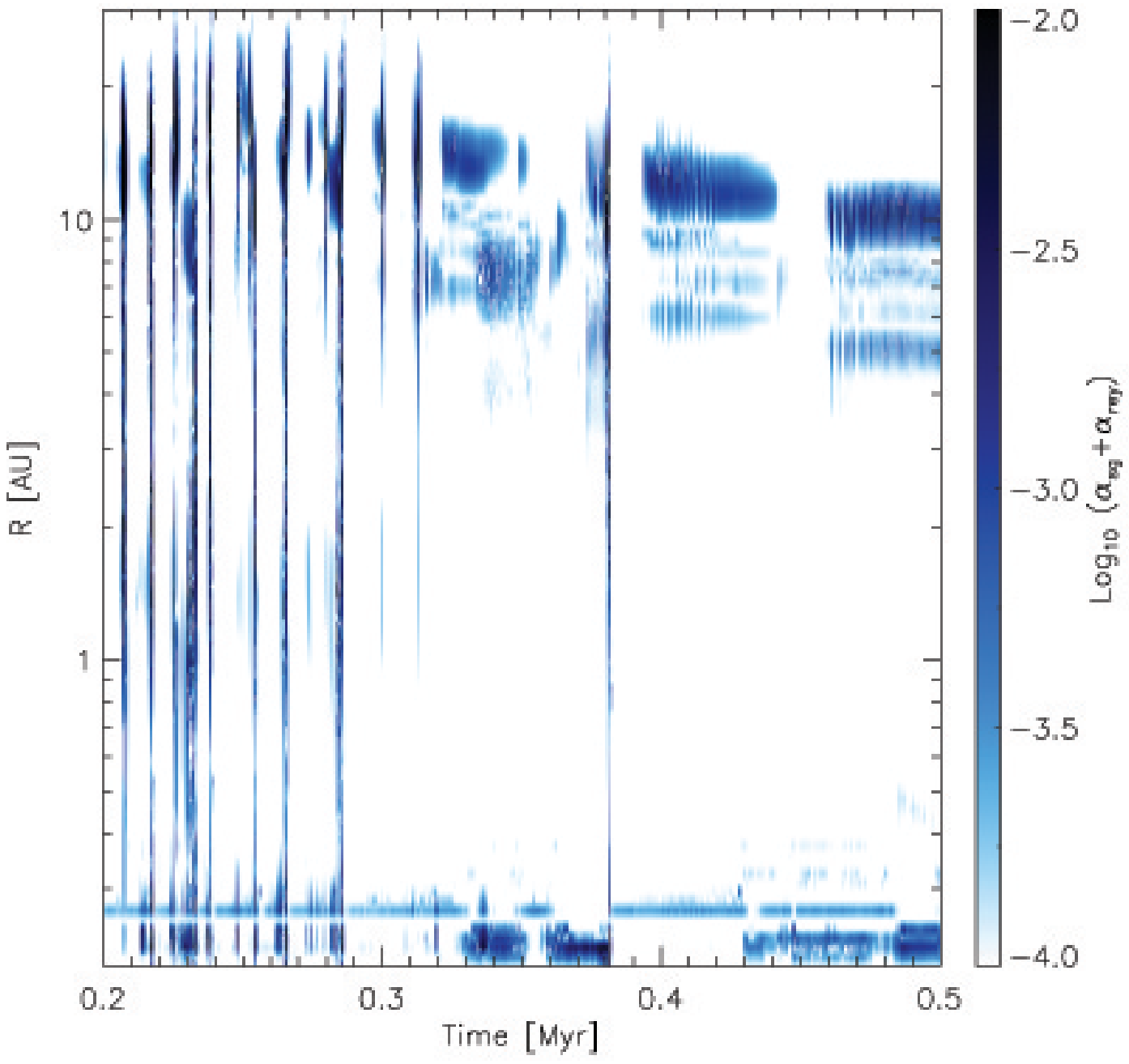}
\caption{Radial profiles of (left) the stress generated directly from the self-gravitating acceleration field $\alpha_{\rm sg}$ and (right) the total stress induced by self-gravity $\alpha_{\rm sg} + \alpha_{\rm rey}$ in logarithmic scale during $t=0.2-0.5$~Myr for the self-gravity model with zero $\alpha_{\rm rd}$. Red triangles on the left panel indicates the time at which outbursts are occurred.}
\label{fig:alphasg}
\end{figure*}

In Figure \ref{fig:alphasg}, we present the time variation of the radial $\alpha_{\rm sg}$ and $\alpha_{\rm rey}$ profiles in the standard self-gravity model over $t=0.2-0.5$~Myr.
We emphasize that the disk repeatedly produces GI-induced stresses which are not constant over time or gradually increasing/decreasing, but are rather sporadic.
This sporadic feature can be understood as a self-regulation process of a disk that stabilizes itself by redistributing mass through the action of spiral waves.  

In terms of the generic $\alpha$ viscosity, this study shows that the total stress driven by GI, while it is a function of time and radius, becomes as large as $\sim0.01$ locally.
This is comparable to the previously used $\alpha$ treatments of disk self-gravity, where an $\alpha_{\rm GI}$ of $0.01-0.03$ \citep{lin87,lin90,armitage01,zhu10a,zhu10b,martin11,martin12,bae13a} is locally assumed for a gravitationally unstable disk region with $Q=1$. 

\subsection{Accretion Outbursts as a Potential Solution to the Luminosity Problem}

\begin{figure*}
\centering
\epsscale{1.1}
\plotone{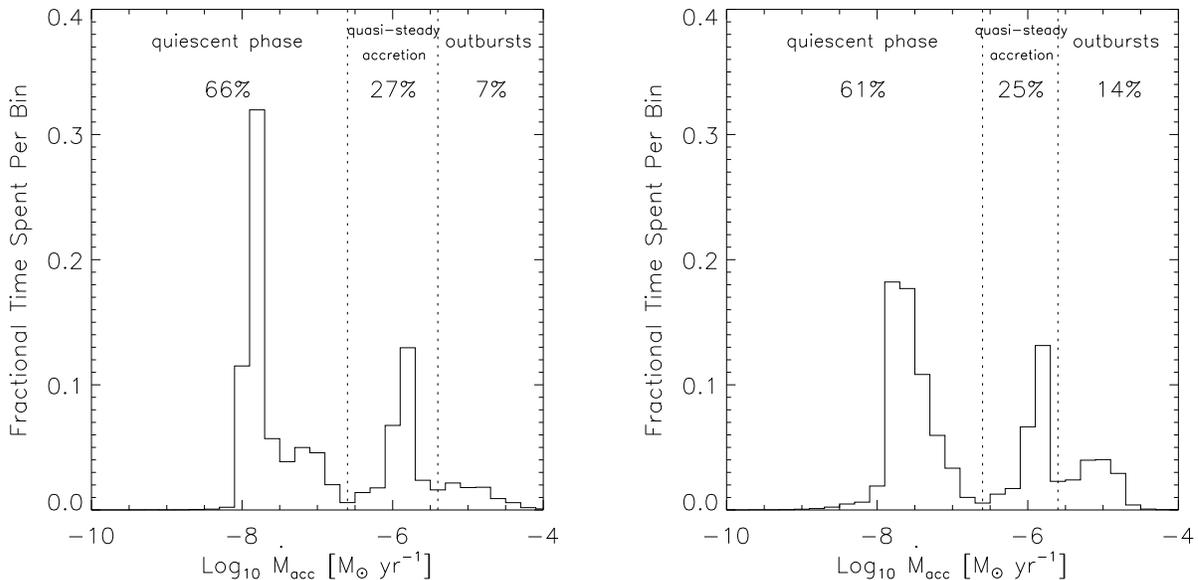}
\caption{Distributions of mass accretion rate during the infall phase with logarithmic bins for self-gravity models with (left) zero and (right) non-zero $\alpha_{\rm rd}$. The histograms can be divided into three phases as indicated by the vertical dotted lines; the early quasi-steady accretion phase, outbursts, and quiescent phase in between bursts. The percentages show fractional time spent in each phase.\\}
\label{fig:luminosity}
\end{figure*}

As mentioned in the Introduction, time-variable protostellar accretion might help 
resolve the luminosity problem in low-mass star formation.
To address the implications of our calculations, in Figure \ref{fig:luminosity} 
we plot the fractional distributions of the mass accretion rate for the infall phase, 
during which time the central protostar is still embedded. 
The highest peak at $\sim10^{-8}~\msunyr$ represents the quiescent disk accretion phase in between outbursts;
this accounts for roughly two-thirds of the total time during infall;
the peak at $\sim10^{-6}~\msunyr$ corresponds to the early phase of quasi-steady disk accretion
at the singular isothermal sphere infall rate, corresponding to about one quarter of the protostellar
phase; and the broad peak at $\gtrsim 10^{-5}~\msunyr$ is due to outbursts, which corresponds to
about $7~\%$ of the infall phase in the zero $\alpha_{\rm rd}$ and about $14~\%$ of the time in
the non-zero $\alpha_{\rm rd}$ model.  For typical mass-radius relations, accretion at $\lesssim 10^{-7}
\msunyr$ produces low enough luminosities to be compatible with observations \citep{kenyon90,offner11}.

While our models illustrate the possibility of outburst behavior to help resolve the luminosity problem by having
protostars spend most of the infall phase accreting slowly, a real test would require constructing a 
luminosity function for an entire population of protostars weighted by the stellar mass function
\citep[e.g.,][]{offner11,dunham12}.  In addition, the distribution of initial angular momenta among the different
mass protostellar clouds would be an important parameter.  
The quasi-steady accretion phase, where infall to the inner disk produces high enough temperatures for the MRI to be activated and thus the disk
accretes at roughly the same rate as the matter falls onto the disk, can be problematic if it persists
for too large a fraction of the infall phase.  In turn, the fraction of time spent in the quasi-steady
phase is a function of the initial angular momentum, because slower rotation leads to more mass being
accreted at small disk radii.  Conversely, large initial angular momenta produce large disks with
accretion strongly modulated by outbursts, as in the models of \citet{vb05,vb06,vb10}.  Further progress
on this problem would be strongly aided by observational constraints on the angular momentum distributions
among protostellar cores of differing masses.

We note that our models, as in those of \citet{zhu10b}, also exhibit outbursts in the post-infall or
T Tauri phase, for which there is little observational evidence.  The mechanisms producing outbursts
in the models are sensitive to the amount of radiative trapping of dissipated energy, which thus
depends upon the surface density and dust opacity; lowering either of these makes it much more difficult
to trigger outbursts.  Thus, over T Tauri lifetimes, removal of mass by photoevaporation
\citep[e.g.,][]{owen11} and dust growth \citep[e.g.][]{miotello14} can reduce the disk opacity and thus radiative 
trapping of thermal energy in the disk becomes less efficient,
lessening the number of outbursts or even preventing them all together.

\subsection{Comments on Other Possible Outburst-Driving Mechanisms}

Thermal instability was one of the first proposed mechanisms aiming to explain the accretion outbursts of FU Ori \citep[e.g.][]{bell94}.
The basic idea is that disk opacity steeply increases between $\sim 2000$~K and few $10^4$~K due to the ionization of hydrogen.
However, raising the disk temperature to such high values to initiate thermal instability is limited only to small radii (few $R_\odot$).
\citet{zhu07} used radiative transfer modeling of FU Ori and found hot inner disk must extend out to $\sim1$~AU, concluding the fit is inconsistent with a pure thermal instability model.
Therefore, while the thermal instability model should not be completely ruled out, we conjecture the model seems to work better when
combined with other mechanisms rather than in isolation.

\citet{vb05,vb06,vb10} suggest that outer disks can fragment and form dense clumps which then migrate inward and eventually accrete onto the central star.
\citet{vb10} included the effect of radiative cooling, viscous and shock heating, stellar and background irradiation and solve disk self-gravity to study protostellar evolution starting from the initial collapse phase.
They found disks fragment at several tens to hundreds AU, whereas we do not see any disk fragmentation in our calculations.
We conjecture this is mainly attributable to the different initial angular momenta assumed in the models.
In terms of angular velocity of collapsing core, this study used $\Omega_c = 1.15 \times 10^{14}~{\rm rad}~{\rm s}^{-1}$ which is the median value inferred by \citet{bae13b}, who reproduced observed disk frequencies as a function of age where disk dispersal by photoevaporation is assumed.
In contrast, the reference model of \citet{vb10} assumed $\Omega_c \sim 9 \times 10^{14}~{\rm rad}~{\rm s}^{-1}$, which is about an order of magnitude greater than ours.
It is also worth to compare the ratio of rotational to gravitational energy $\beta = E_{\rm rot}/|E_{\rm grav}|$.
In this study, we use a two-component density profile for the initial Bonnor-Ebert sphere which is described as
\be
\rho = \rho_c~{\rm at}~ \xi < \xi_c
\en
and
\be
\rho = 2\rho_c \xi^{-2}~{\rm at}~ \xi_c < \xi < 6.5,
\en
where $\rho_c$ is the central density and $\xi = r/(c_s^2/4\pi G \rho_c)^{1/2}$ is the non-dimensional radial distance.
Note that the density profile beyond $\xi = \xi_c$ has the same profile as the singular isothermal model, and $\xi = 6.5$ corresponds to the critical Bonnor-Ebert sphere radius.
As our initial conditions assume the flat, inner part of the Bonnor-Ebert is collapsed to $0.2~\msun$ central protostar leaving outer $1~\msun$ of envelope cloud, the corresponding $\xi_c$ becomes 1.78.
With this initial setup $\beta=3.0\times 10^{-4}$, which is smaller than the one used in the reference model of \citet{vb10} by a factor of $\sim40$.

We also note that while the suggested process in \citet{vb05,vb06,vb10} seems plausible, it is uncertain whether the clumps created at relatively large radii eventually accrete onto the central star and lead to a rise in the accretion rate given their placement of the inner boundary at a relatively large
radius ($R_{\rm in}$ = 5~AU).
For instance, it may be possible that the clumps are tidally destroyed as they migrate \citep{zhu12}.
With such a large inner boundary one can also miss important physics including GI + MRI and thermal triggering of outbursts at smaller radii as we show in this paper.
We tested our model with an inner boundary of $R_{\rm in}$ = 5~AU and not surprisingly found that neither GI + MRI-driven nor thermally driven outbursts occur.

\section{CONCLUSIONS}

In this paper, we explicitly solve disk self-gravity to investigate the triggering of accretion outbursts in two dimension starting from the collapse of an isothermal, uniformly-rotating core.
We find that gravitationally unstable disks generate spiral density waves that heat disks via compressional heating and can trigger accretion outbursts by activating the MRI in the disk dead-zone.
We emphasize that the GI-induced spiral waves can propagate well inside of the gravitationally unstable region 
before they trigger outbursts at $R \lesssim 1$~AU; this feature cannot be reproduced with the previously 
used local $\alpha_{\rm GI}$ treatments.  
As suggested in our previous one-dimensional calculations (Paper I), we further confirm that the presence of a small but finite $\alpha_{\rm rd}$ of $10^{-4}$ triggers thermally-driven bursts of accretion soon after mass feeding from envelope cloud is ceased, instead of GI + MRI-driven outbursts.
We argue that the episodic mass accretion during protostellar evolution can qualitatively help explain the 
low accretion luminosities seen in low-mass protostars, while allowing the protostars to grow in mass on the requisite time scales,
although a proper test will require calculations
for differing final protostellar masses as well as some constraint on the distribution of angular momenta as a function
of protostellar core mass. 

Our current models include only a very crude treatment of the activation of the MRI, and this can strongly affect the
detailed nature of the outbursts in the inner disk.  Better predictions of accretion luminosities will require three-dimensional
magnetohydrodynamic simulations which can treat the MRI activation in the innermost disk.

\acknowledgments

This work was supported in part by NADA grant NNX11AK53G, and computational resources and services provided by Advanced Research Computing at the University of Michigan, Ann Arbor.
Z.Z. acknowledges support by NASA through Hubble Fellowship grant HST-HF-51333.01-A awarded by the Space Telescope Science Institute, which is operated by the Association of Universities for Research in Astronomy, Inc., for NASA, under contract NAS 5-26555.

\appendix

\section{Infall Heating}
\label{app:infall}

Here, we derive the infall heating by shock dissipation given in Equation (\ref{eqn:q_infall}).
Assuming an axisymmetric infall model for simplicity, mass, angular momentum, and energy equations in cylindrical coordinates are
\be
\label{eqn:mass_1d}
R {\partial \Sigma \over \partial t} - {1 \over 2\pi} {\partial \dot{M} \over \partial R} = R \dot{\Sigma}_{\rm in},
\en
\be
\label{eqn:ang_momentum_1d}
R {\partial\over \partial t}(\Sigma R^2 \Omega) - {1 \over 2 \pi} {\partial \over \partial R}(\dot{M}  R^2 \Omega) = {\partial \over \partial R} (R^2 \Pi_{R\phi}) + R^2 \dot{\Sigma}_{\rm in} v_{\phi,{\rm in}},
\en
and
\be
\label{eqn:energy_1d}
R {\partial \over \partial t}(\Sigma E) - {1 \over 2 \pi} {\partial \over \partial R}(\dot{M} E) = R Q_+ - 2 R \sigma T^4.
\en
In the above equations $\Sigma$ is the surface density, $\dot{M}$ is the radial mass flux defined as $\dot{M} \equiv - 2 \pi R \Sigma v_R$, $\dot{\Sigma}_{\rm in}$ is the infall rate defined as $\dot{\Sigma}_{\rm in}=\dot{M}_{\rm in}/2\pi R_c R$, $\Omega$ is the angular velocity, $\Pi_{R\phi}$ is $R-\phi$ component of the viscous stress tensor, $M_*$ is the stellar mass, $R_c$ is the centrifugal radius, $E$ is the total energy per unit mass except thermal energy, $Q_+$ includes all heating sources except the infall heating, and $T$ is the disk temperature.
Assuming instantaneous centrifugal balance and $\Pi_{R\phi}=R\Sigma \nu d\Omega/dR$, Equations (\ref{eqn:mass_1d}) and (\ref{eqn:ang_momentum_1d}) can be simplified to 
\be
\label{eqn:mdot_1d}
\dot{M} = 6\pi R^{1/2} {\partial \over \partial R} (R^{1/2} \Sigma \nu) + {2\pi R^2 \Sigma \over M_*} {\partial M_* \over \partial t} - 4\pi R^2 \dot{\Sigma}_{\rm in} \left[ \left({R \over R_c}\right)^{1/2} -1 \right].
\en
For the next, combining Equations (\ref{eqn:mass_1d}) and (\ref{eqn:energy_1d}) gives
\be
\label{eqn:q_infall_1d}
\Sigma {\partial E\over \partial t} = Q_+ - \dot{\Sigma}_{\rm in} E + {\dot{M} \over {2\pi R}}{\partial E \over \partial R} - 2\sigma T^4.
\en
From now on, let us focus on the terms induced from infall only.
By substituting $\dot{M}$ in Equation (\ref{eqn:q_infall_1d}) with Equation (\ref{eqn:mdot_1d}) we get the total heating due to infall as follows.
\be
\label{eqn:infall_heating}
 Q_{\rm in, total}  = -\dot{\Sigma}_{\rm in} E - 2 R  \dot{\Sigma}_{\rm in} \left[ \left({R \over R_c}\right)^{1/2} -1  \right] {\partial E \over \partial R}
\en

\begin{figure}
\centering
\epsscale{0.6}
\plotone{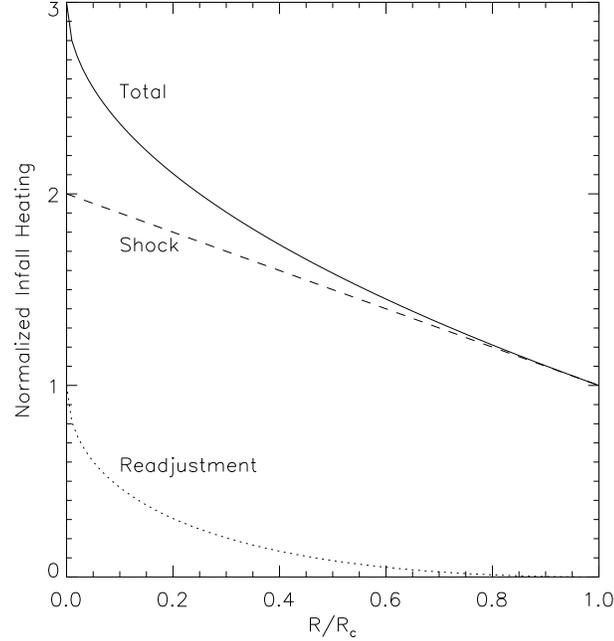}
\caption{Normalized infall heating as a function of radius. The total infall heating is plotted with a solid curve while heating through the instantaneous shock dissipation and the readjustment process are plotted with a dashed and a dotted curve, respectively. At the centrifugal radius $R_c$ infalling material arrives at the disk surface nearly horizontally with the Keplerian azimuthal velocity, so all the kinetic energy is dissipated through shocks.}
\label{fig:infall}
\end{figure}

When it arrives at the disk surface infalling material has velocity of 
\begin{eqnarray}
v_R & = & - \left( {GM_* \over R} \right)^{1/2}
\\
v_\theta & = & \left({GM_* \over R}\right)^{1/2} \cos \theta_0
\\
v_\phi & = & \left( {GM_* \over R} \right)^{1/2}\sin \theta_0
\end{eqnarray}
where $\theta_0$ is the angle between the orbital plane and the rotation axis of the system and $\sin^2 \theta_0 = R/R_c$ at the disk surface \citep{cassen81}.
Thus, infalling material brings zero total energy ($E_{\rm tot} = E_{\rm kin} + E_{\rm pot} = GM_*/R - GM_*/ R = 0$), while disk material has total energy of $-GM_*/2R$ assuming a Keplerian disk.
Using Equation (\ref{eqn:infall_heating}) the total infall heating that corresponds to the additional energy of infalling material is 
\be
Q_{{\rm in,total}} = {GM_* \dot{M}_{\rm in} \over 4\pi R_c^3} {3-2 (R/R_c)^{1/2} \over (R/R_c)^2}.
\en
At the disk surface, only the kinetic energy corresponding to the $v_R$ and $v_\theta$ component of the infall is released instantaneously through the shock, which is $(2-R/R_c)GM_*/2R$.
The heat dissipated through the shock dissipation is then
\be
Q_{{\rm in,shock}} = {GM_* \dot{M}_{\rm in} \over 4\pi R_c^3} {2-(R/R_c) \over (R/R_c)^2}.
\en
The rest of the additional energy is taken care by the code with a proper shear force term in the momentum equation, which would correspond to
\be
Q_{{\rm in,readjust}} = {GM_* \dot{M}_{\rm in} \over 4\pi R_c^3} {1 + (R/R_c) - 2 (R/R_c)^{1/2} \over (R/R_c)^2}.
\en
The normalized infall heating profile as a function of radius is presented in Figure \ref{fig:infall} to show their relative importance at each radius.

\end{document}